\documentclass[12pt]{iopart}
\usepackage{iopams}
\usepackage{setstack}
\usepackage{psfig}

\newcommand{\cP}{\mathcal P}

\newcommand{\cT}{\mathcal T}
\begin{document}

\title[Classical Trajectories for Complex Hamiltonians]
{Classical Trajectories for Complex Hamiltonians}

\author[Bender, Chen, Darg, and Milton]{Carl~M~Bender$^*$, Jun-Hua Chen$^*$,
Daniel W. Darg$^\dag$, and Kimball A. Milton$^*$\footnote{Permanent address:
Homer L.~Dodge Department of Physics and Astronomy,
University of Oklahoma, Norman, OK 73019, USA}}

\address{${}^*$Department of Physics, Washington University, St. Louis MO
63130, USA}

\address{${}^\dag$Blackett Laboratory, Imperial College, London SW7 2BZ, UK}

\begin{abstract}
It has been found that complex non-Hermitian quantum-mechanical Hamiltonians may
have entirely real spectra and generate unitary time evolution if they possess
an unbroken $\cP\cT$ symmetry. A well-studied class of such Hamiltonians is $H=
p^2+x^2(ix)^\epsilon$ ($\epsilon\geq0$). This paper examines the underlying
classical theory. Specifically, it explores the possible trajectories of a
classical particle that is governed by this class of Hamiltonians. These
trajectories exhibit an extraordinarily rich and elaborate structure that
depends sensitively on the value of the parameter $\epsilon$ and on the initial
conditions. A system for classifying complex orbits is presented.
\end{abstract}

% (\today)

\submitto{\JPA}

%\maketitle

\section{Introduction}
\label{s1}

There are huge classes of complex $\cP\cT$-symmetric non-Hermitian
quantum-mechanical Hamiltonians whose spectra are real and which exhibit unitary
time evolution. A particularly interesting class of such Hamiltonians is
\cite{r1,r2,r3}
\begin{equation}
H=p^2+x^2(ix)^\epsilon\qquad(\epsilon\geq0).
\label{e1}
\end{equation}
An almost obvious question to ask is, What is the nature of the underlying
classical theory described by this Hamiltonian?

This question was addressed in several previous studies \cite{r4,r5}. These
papers presented numerical studies of the classical trajectories, that is, the
position $x(t)$ of a particle of a given energy as a function of time. Some
interesting features of these trajectories were discovered:
\begin{itemize}
\item While $x(t)$ for a Hermitian Hamiltonian is a real function, a complex
Hamiltonian typically generates complex classical trajectories. Thus, even if
the classical particle is initially on the real-$x$ axis, it is subject to
complex forces and thus it will move off the real axis and travel through the
complex plane.

\item For the Hamiltonian in (\ref{e1}) the classical domain is a multisheeted
Riemann surface when $\epsilon$ is noninteger. In this case, the classical
trajectory may visit more than one sheet of the Riemann surface. Indeed, in
Ref.~\cite{r4} classical trajectories that visit three sheets of the Riemann
surface were displayed.

\item Because $\epsilon\geq0$, the $\cP\cT$ symmetry of $H$ in (\ref{e1}) is
unbroken \cite{r3} and, as a result, the classical orbits are closed periodic
paths in the complex plane. When $\epsilon$ is negative, the classical
trajectories are open (and nonperiodic).

\item The classical trajectories manifest the $\cP\cT$ symmetry of the
Hamiltonian. Under parity reflection $\cP$ the position of the particle changes
sign: $\cP:\,x(t)\to-x(t)$. Under time reversal $\cT$ the sign of both $t$ and
$i$ are reversed, so $\cT:\,x(t)\to x^*(-t)$. Thus, under combined $\cP\cT$
reflection the classical trajectory is replaced by its mirror image with
respect to the imaginary axis on the principal sheet of the Riemann surface.
\end{itemize}

Although these features of classical non-Hermitian $\cP\cT$-symmetric
Hamiltonians were already known, we show in this paper that
the structure of the complex trajectories is
much richer and more elaborate than was previously noticed. One can find
trajectories that visit huge numbers of sheets of the Riemann surface and
exhibit fine structure that is exquisitely sensitive to the initial condition
$x(0)$ and to the value of $\epsilon$. Small variations in $x(0)$ and $\epsilon$
give rise to dramatic changes in the topology of the classical orbits and to
the size of the period. We show in Sec.~\ref{s2} that depending on the value of
$x(0)$ there are periodic orbits having short periods as well as orbits having
extremely long and possibly even infinitely long periods. These results are
reminiscent of the period-lengthening route to chaos that is observed in
logistic maps \cite{r6}. The period of a classical orbit is discussed in
Sec.~\ref{s3}, where we show that the period depends on the topology of the
orbit. In particular, the period depends on the specific pairs of turning points
that are enclosed by the orbit and on the number of times that the orbit
encircles each pair. We use the period to characterize the topology of the
orbits. For a given initial condition the classical behavior undergoes
remarkable transitions as $\epsilon$ is varied. There are narrow regions at
whose boundaries we observe critical behavior in the topology of the classical
orbits as well as large regions of quiet stability. This striking dependence on
$\epsilon$ is elucidated in Sec.~\ref{s4}. Finally, in Sec.~\ref{s5} we make
some concluding observations.

\section{Dependence of classical orbits on initial conditions}
\label{s2}

In this section we study the dependence on initial conditions
of classical orbits governed by (\ref{e1}).
To construct the classical trajectories, we first note that the
value of the Hamiltonian in (\ref{e1}) is a constant of the motion. Without loss
of generality, this constant (the energy $E$) may be chosen to be 1. (If $E$
were not 1, we could then rescale $x$ and $t$ to make $E=1$.) Because $p(t)$ is
the time derivative of $x(t)$, the trajectory $x(t)$ satisfies a first-order
differential equation whose solution is determined by the initial condition
$x(0)$ and the sign of $\dot x(0)$.

\begin{figure*}[t!]
\vspace{2.4in}
\includegraphics{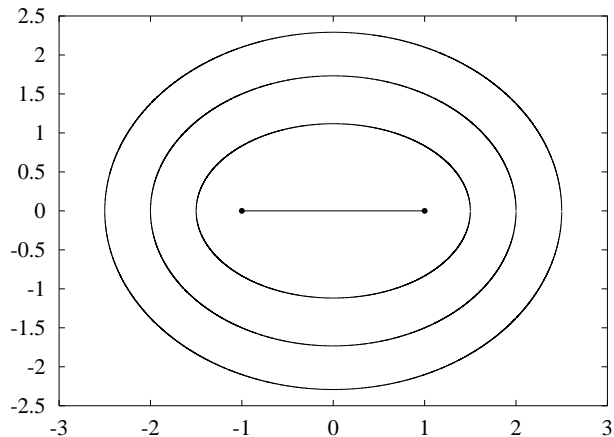}
\caption{Classical trajectories in the complex-$x$ plane for the harmonic
oscillator whose Hamiltonian is $H=p^2+x^2$. These trajectories represent the
possible paths of a particle whose energy is $E=1$. The trajectories are nested
ellipses with foci located at the turning points at $x=\pm1$. The real line
segment (degenerate ellipse) connecting the turning points is the usual periodic
classical solution to the harmonic oscillator. All closed paths have the same
period $\pi$ by virtue of Cauchy's integral theorem.}
\label{f1}
\end{figure*}

Let us begin by examining the harmonic oscillator, which is obtained by setting
$\epsilon=0$ in (\ref{e1}). For the harmonic oscillator the turning points (the
solutions to the equation $x^2=1$) lie at $x=\pm1$. If we chose $x(0)$ to lie
between these turning points,
\begin{equation}
-1\leq x(0)\leq1,
\label{e2}
\end{equation}
then the classical trajectory oscillates between the turning points with period
$\pi$. This orbit is shown in Fig.~\ref{f1} as the solid horizontal line
joining the turning points.

However, while the harmonic-oscillator Hamiltonian is Hermitian, it can still
have complex classical trajectories. To obtain one of these trajectories, we
choose an initial condition that does not lie between the turning points and
thus does not satisfy (\ref{e2}). The resulting trajectories are ellipses in the
complex plane (see Fig.~\ref{f1}). The foci of these ellipses are the turning
points \cite{r4}. Note that for each of these closed orbits the period is always
$\pi$; this is a consequence of the Cauchy integral theorem applied to the
integral that represents the period.

\begin{figure*}[t!]
\vspace{2.4in}
\includegraphics{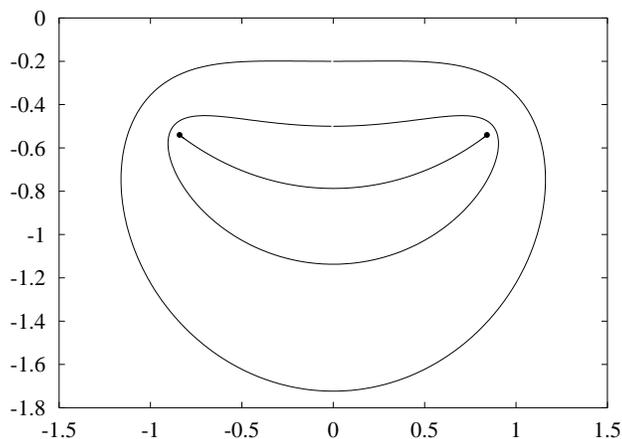}
\caption{Classical trajectories in the complex-$x$ plane for the complex
oscillator whose Hamiltonian is $H=p^2-(ix)^\pi$, which is (\ref{e1}) with
$\epsilon=\pi-2$. As in
Fig.~\ref{f1} the trajectories represent the possible paths of a particle whose
energy is $E=1$. The trajectories are deformed versions of the ellipses in
Fig.~\ref{f1}. By virtue of Cauchy's integral theorem all of the closed
trajectories have the same period $T$ as given in (\ref{e5}).}
\label{f2}
\end{figure*}

As $\epsilon$ increases from 0, the pair of turning points at $x=\pm1$ moves
downward into the complex-$x$ plane. These turning points are determined by the
equation
\begin{equation}
1+(ix)^{2+\epsilon}=0.
\label{e3}
\end{equation}
When $\epsilon$ is noninteger, this equation has many solutions, all having
absolute value 1. These solutions have the form
\begin{equation}
x=\exp\left(i\pi\frac{4N-4-\epsilon}{4+2\epsilon}\right),
\label{e4}
\end{equation}
where $N$ is an integer. These turning points occur in $\cP\cT$-symmetric pairs
(that is, pairs that are reflected through the imaginary axis) corresponding to
the $N$ values $(N=1,~N=0)$, $(N=2,~N=-1)$, $(N=3,~N=-2)$, $(N=4,~N=-3)$, and
so on. We label these pairs by the integer $n$ ($n=0,~1,~2,~3,~\ldots$) so that
the $n$th pair corresponds to $(N=n+1,~N=-n)$.
Note that the pair of turning points at $\epsilon=0$
deforms continuously into the $n=0$ pair of turning points when $\epsilon\neq0$.
For the case $\epsilon=\pi-2$
these turning points are shown in Fig.~\ref{f2} as dots.

In Fig.~\ref{f2} three closed classical trajectories are shown. First, there is 
the path connecting the $n=0$ turning points, which is a deformed version of the
straight line in Fig.~\ref{f1}. Two other trajectories that enclose
these two turning points
are also indicated. These closed orbits are deformations of the ellipses
shown in Fig.~\ref{f1}. Furthermore, as in the $\epsilon=0$ case, the Cauchy
integral theorem implies that the period $T$ for each of these orbits is the
same. The general formula for the period of a closed orbit whose topology is
like that of the orbits shown in Fig.~\ref{f2} is
\begin{eqnarray}
T=2\sqrt{\pi}{\Gamma\left({3+\epsilon\over2+
\epsilon}\right)\over\Gamma\left({4+\epsilon\over4+2\epsilon}\right)}
\cos\left({\epsilon\pi\over4+2\epsilon}\right).
\label{e5}
\end{eqnarray}
This formula is given in Ref.~\cite{r4} and is valid for all $\epsilon\geq0$.
For the case of the closed orbits shown in Fig.~\ref{f2}, we find that
$T=2.33276$.

The derivation of (\ref{e5}) is straightforward. The period $T$ is given by a
closed contour integral along the trajectory in the complex-$x$ plane. This
trajectory encloses the square-root branch cut that joins the turning points.
This contour can be deformed into a pair of rays that run from one turning point
to the origin and then from the origin to the other turning point. The integral
along each ray is easily evaluated as a beta function, which is then written in
terms of gamma functions.

\begin{figure*}[t!]
\vspace{2.4in}
\includegraphics{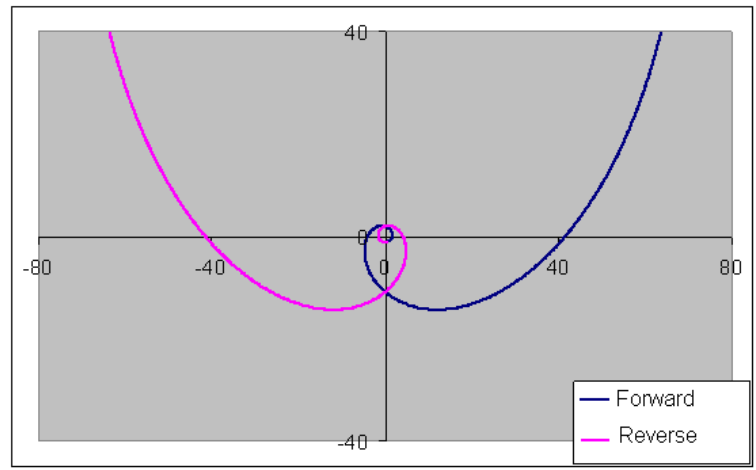}
\caption{Classical trajectories in the complex-$x$ plane for the Hamiltonian in
(\ref{e1}) with $\epsilon=-0.2$. These trajectories begin on the negative
imaginary axis very close to the origin. One trajectory evolves forward in time
and the other goes backward in time. The trajectories are open orbits and show
the particle spiraling off to infinity. The trajectories begin on the principal
sheet of the Riemann surface; as they cross the branch cut on the
positive imaginary axis, they visit the higher and lower sheets of the surface.
Note that the trajectories do not cross because they lie on different sheets.}
\label{f3}
\end{figure*}

The key difference between classical paths for $\epsilon>0$ and for $\epsilon<0$
is that in the former case all the paths are closed orbits and in the latter
case the paths are open orbits. In Fig.~\ref{f3} we consider the case $\epsilon=
-0.2$ and display two paths that begin on the negative imaginary axis. One path
evolves forward in time and the other path evolves backward in time. Each path
spirals outward and eventually moves off to infinity. Note that the pair of
paths is a $\cP\cT$-symmetric structure. Note also that the paths do not cross
because they are on different sheets of the Riemann surface. The function $(ix)^
{0.2}$ requires a branch cut, and we take this branch cut to lie along the
positive imaginary axis. The forward-evolving path leaves the principal sheet
(sheet 0) of the Riemann surface and crosses the branch cut in the positive
sense and continues on sheet 1. The reverse path crosses the branch cut in the
negative sense and continues on sheet $-1$. Figure \ref{f3} shows the projection
of the classical orbit onto the principal sheet.

\begin{figure*}[t!]
\vspace{2.4in}
\includegraphics{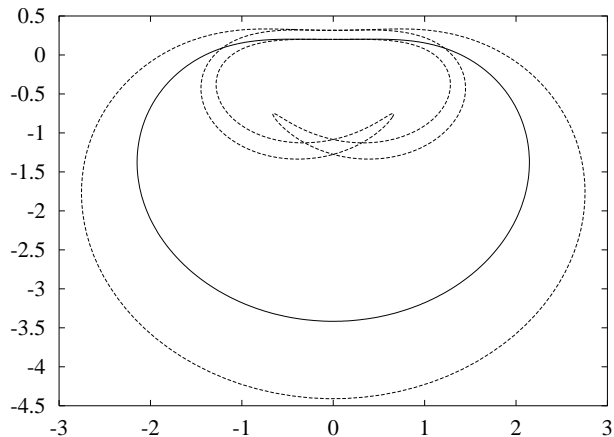}
\caption{A classical trajectory in the complex-$x$ plane for the Hamiltonian
$H=p^2-(ix)^\pi$, which is obtained by setting $\epsilon=\pi-2$ in (\ref{e1}).
The initial condition is chosen so that the path crosses the 
branch cut on the positive imaginary axis and leaves the principal sheet of
the Riemann surface. On the principal sheet the trajectory is indicated by a
solid line. The classical particle visits two other sheets of the Riemann
surface on which the trajectory is indicated by a dashed line. Note that the
closed orbit is $\cP\cT$ symmetric (has left-right symmetry) and that the period
is $T=11.8036$.}
\label{f4}
\end{figure*}

Let us now examine closed orbits having a more complicated topological structure
than the orbits shown in Fig.~\ref{f2}. For the rest of this section we fix
$\epsilon=\pi-2$ and study the effect of varying the initial conditions. It is
not difficult to find an initial condition for which the classical trajectory
crosses the branch cut on the positive imaginary axis and leaves the principal
sheet of the Riemann surface. In Fig.~\ref{f4} we show such a trajectory. This
trajectory visits three sheets of the Riemann surface, the principal sheet
(sheet 0) on which the trajectory is shown as a solid line, and sheets $\pm1$ on
which the trajectory is shown as a dashed line. On the Riemann surface the
resulting trajectory is $\cP\cT$-symmetric (left-right symmetric).

The period of the orbit in Fig.~\ref{f4} is $T=11.8036$, which is roughly five
times longer than the periods of the orbits shown in Fig.~\ref{f2}. This is
because the orbit is topologically more complicated and encloses branch cuts
joining three pairs rather than one pair of complex turning points.
(The period of the orbit is roughly proportional to the number of times that
the orbit crosses the imaginary axis.)
We explain
how to calculate the period of these topologically nontrivial orbits in
Sec.~\ref{s3}.

\begin{figure*}[t!]
\vspace{2.9in}
\includegraphics{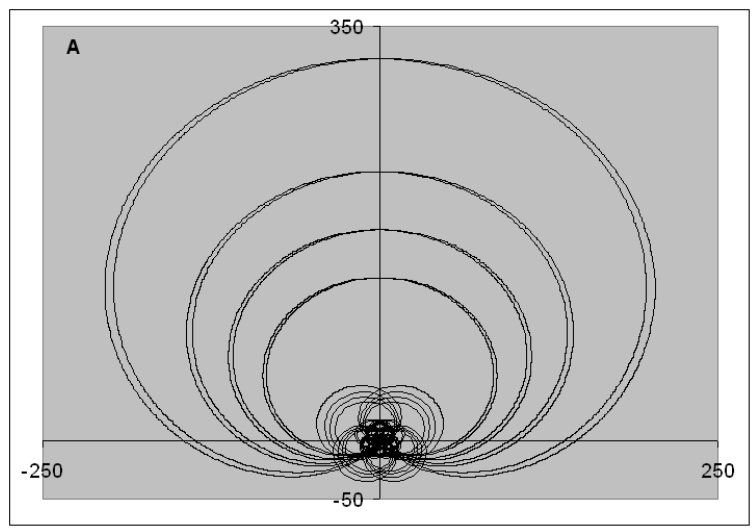}
\caption{A classical trajectory in the complex-$x$ plane for the complex
Hamiltonian $H=p^2-(ix)^\pi$. This complicated trajectory begins at
$x(0)=-7.1i$ and visits 11 sheets
of the Riemann surface. Its period is approximately $T=255.3$.
This figure displays the projection of the trajectory onto the principal
sheet of the Riemann surface. Note that this trajectory does not cross itself.}
\label{f5}
\end{figure*}

\begin{figure*}[t!]
\vspace{2.4in}
\includegraphics{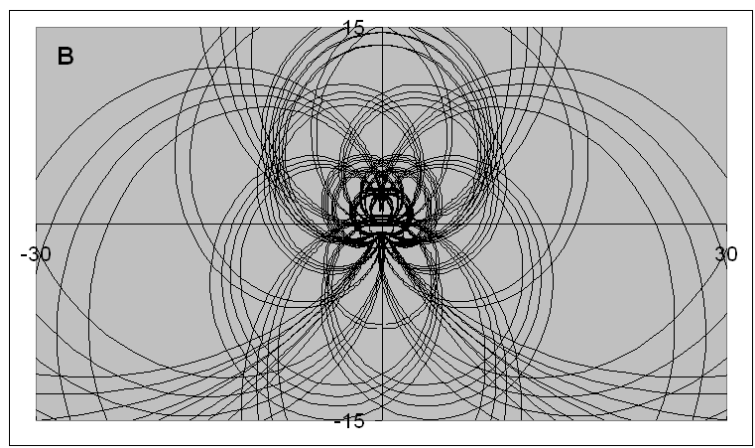}
\caption{An enlargement of the classical trajectory $x(t)$ in Fig.~\ref{f5}
showing the detail near the origin in the complex-$x$ plane. We emphasize
that this classical path never crosses itself; the apparent self-intersections
are paths that lie on different sheets of the Riemann surface.}
\label{f6}
\end{figure*}

The closed orbit shown in Fig.~\ref{f4} only visits three sheets of the
Riemann surface. It is possible to find initial conditions that generate
trajectories that visit many sheets repeatedly. In Fig.~\ref{f5} we have
plotted a classical trajectory starting at $x(0)=-7.1i$. This trajectory visits
11 sheets of the Riemann surface and its period is $T=255.3$. The structure of
this orbit near the origin is complicated and therefore a magnified version is
shown in Fig.~\ref{f6}.

\begin{figure*}[t!]
\vspace{2.6in}
\includegraphics{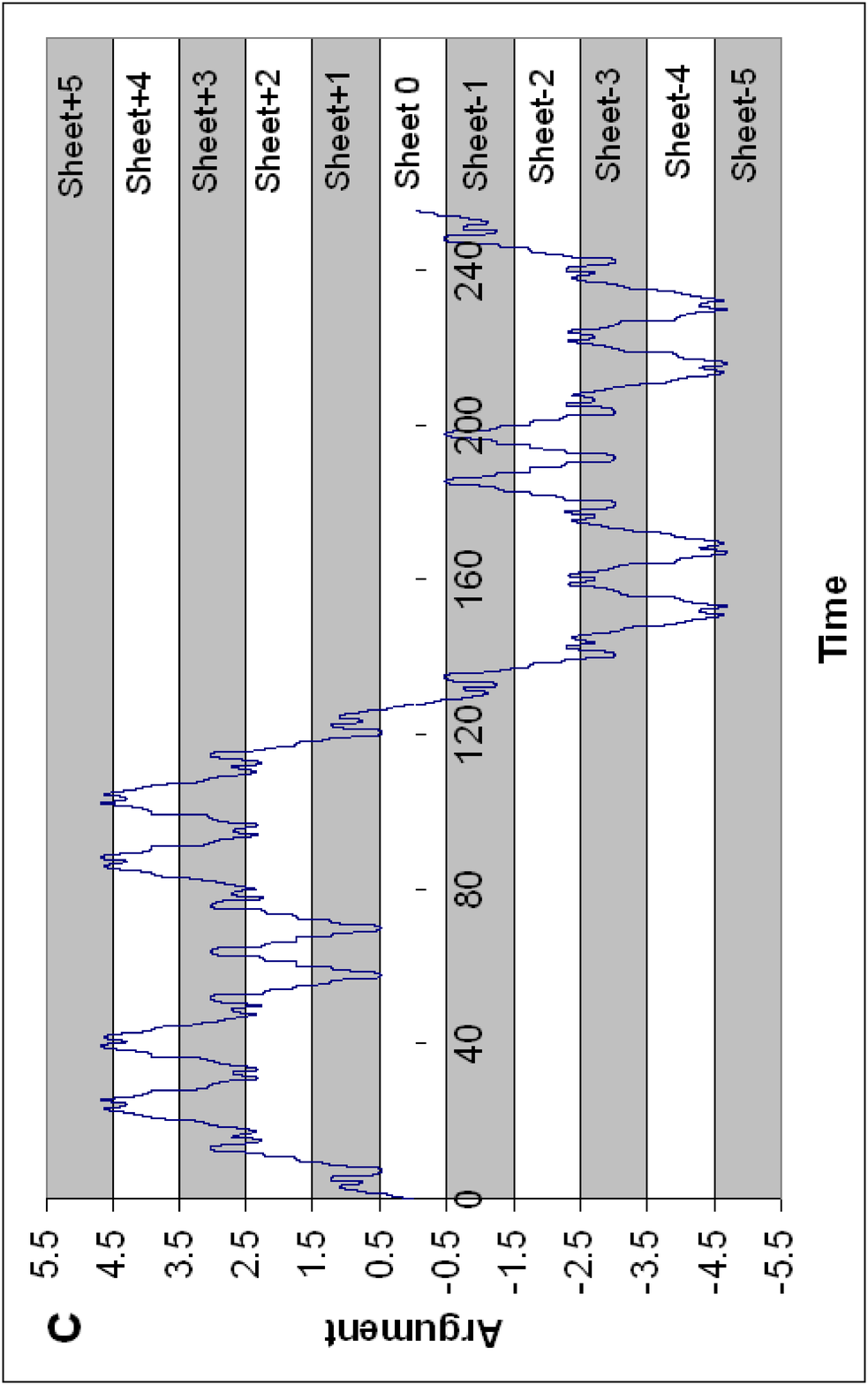}
\caption{The argument (complex phase) of the classical orbit shown in
Figs.~\ref{f5} and \ref{f6} plotted as a function of time for one complete
cycle. The period of this cycle is $T=255.3$. The classical particle
starts on the negative imaginary axis on sheet 0 where the phase is defined
to be 0. The particle then
visits 11 sheets of the Riemann surface from sheet $-5$ to sheet 5.}
\label{f7}
\end{figure*}

Because Figs.~\ref{f5} and \ref{f6} are so complicated, it is useful to give a
more understandable representation of the classical orbit in which we plot the
complex phase (argument) of $x(t)$ as a function of $t$. In Fig.~\ref{f7} we
present such a plot showing the complex phase for one full period.

\begin{figure*}[t!]
\vspace{2.5in}
\includegraphics{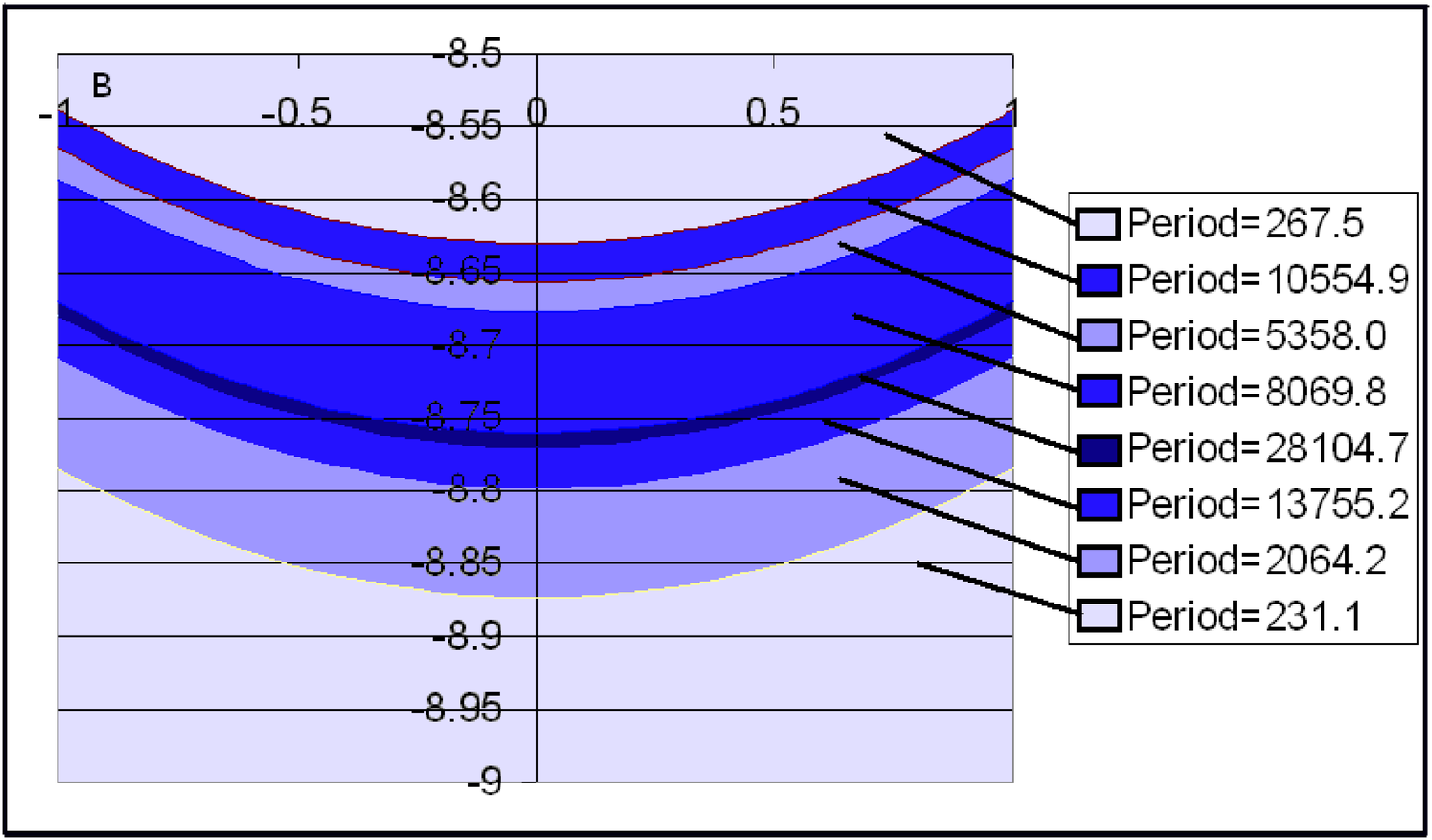}
\caption{A small portion of the complex-$x$ plane showing the dependence of
the periods $T$ of the classical orbits on the choice of initial condition
$x(0)$ for the case $\epsilon=\pi-2$. Note that $T$ is extremely sensitive to 
the value of $x(0)$. There is an unresolved region between the band
corresponding to $T=28,104.7$ and $T=13,755.2$. We conjecture that arbitrarily
long periods can be found in arbitrarily thin regions between $x(0)=-8.767i$
and $x(0)=-8.770i$.}
\label{f8}
\end{figure*}

The period of the classical orbits is exquisitely sensitive to the initial
conditions. To illustrate this sensitivity we show in Fig.~\ref{f8} the size
of the period for $\epsilon=\pi-2$ as a function of the initial condition $x(0)$
in a small
portion of the complex-$x$ plane containing the negative imaginary axis from
$-8.5i$ to $-9.0i$. Note that initial conditions chosen from this small region
give rise to classical orbits whose periods range from $231.1$ up to $28,104.7$.
The regions of extremely long periods become narrower and more difficult to
observe numerically. It is impossible to resolve the fine detail
between the two longest periods, and we conjecture that there are infinitely
many arbitrarily thin regions of initial conditions between $-8.767i$ and
$-8.770i$ that give rise to arbitrarily long periods.

\begin{figure*}[t!]
\vspace{2.5in}
\includegraphics{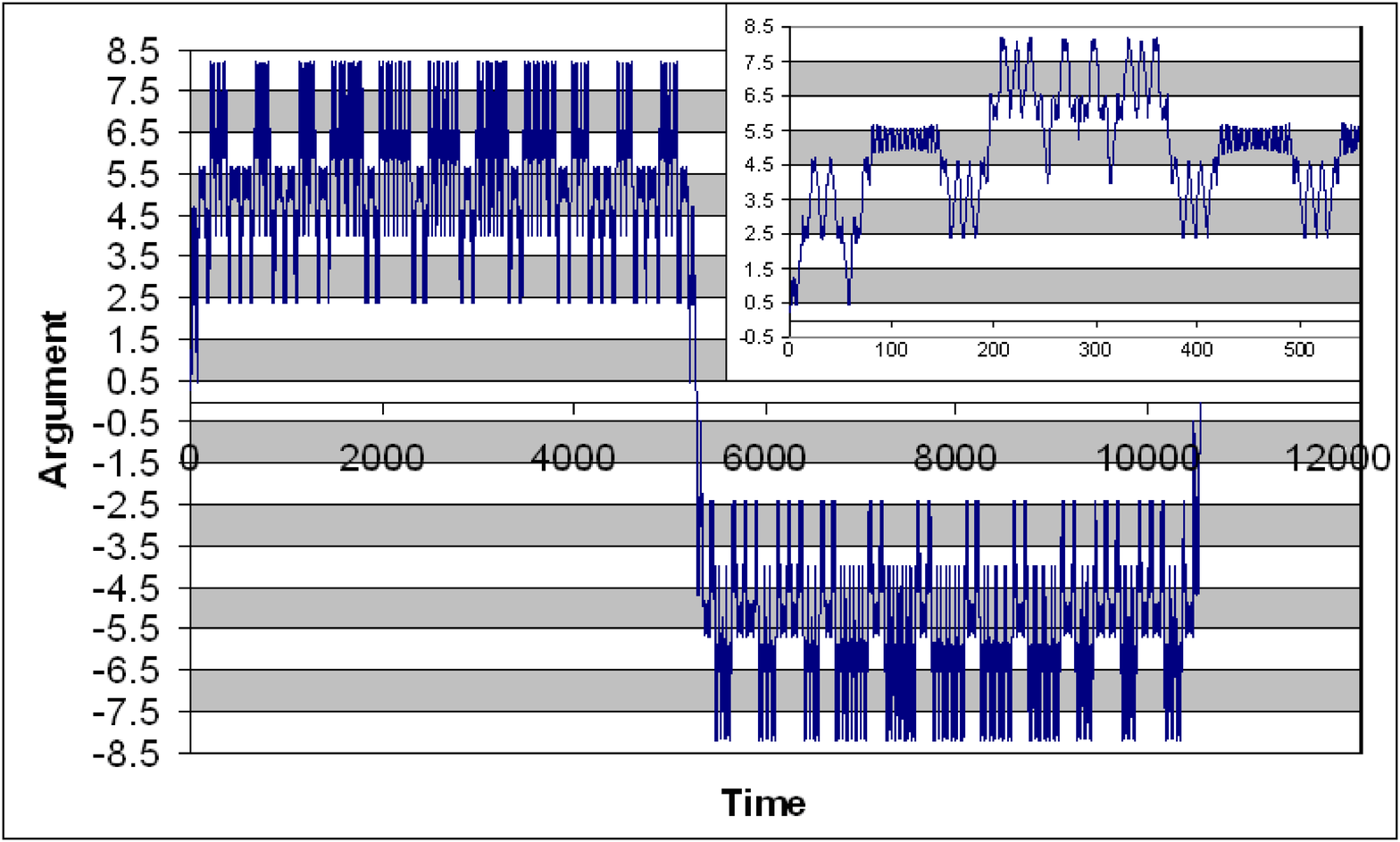}
\caption{The argument of a long-period classical orbit for which 
$\epsilon=\pi-2$ and the initial condition is $x(0)=-8.63026i$.
This orbit has
period $T=10,554.9$ and visits 17 sheets of the Riemann surface.
Note the oscillatory fine structure of this orbit in the inset.}
\label{f9}
\end{figure*}

We display in Fig.~\ref{f9} one of the long-period orbits taken from
Fig.~\ref{f8}. Figure \ref{f9} shows
the complex argument of $x(t)$ as a function of time $t$ for
$\epsilon=\pi-2$ and initial condition $x(0)=-8.63026i$. This orbit has
period $T=10,554.9$ and visits 17 sheets of the Riemann surface.
The inset displays some of the fine structure of this spectacular
oscillatory behavior.

\begin{figure*}[t!]
\vspace{2.6in}
\includegraphics{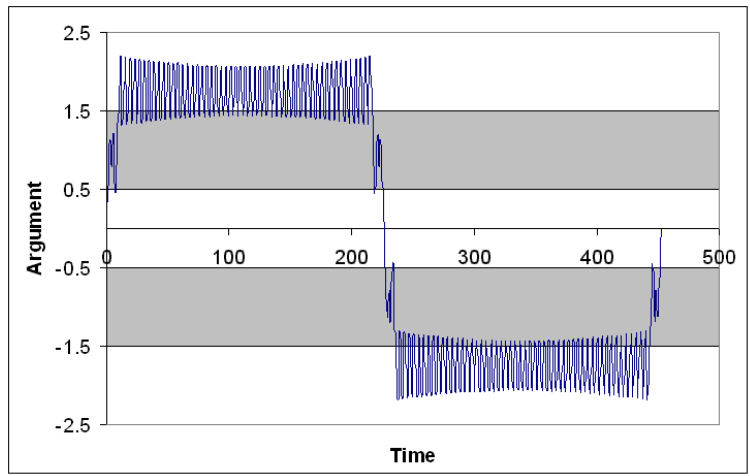}
\caption{The argument of the classical orbit as a function of time $t$ for
$\epsilon=\pi-2$ and initial condition $x(0)=-17i$. This orbit has
period $T=452.6$ and visits 5 sheets of the Riemann surface.
Note the persistent oscillation in the classical orbit.}
\label{f10}
\end{figure*}

\begin{figure*}[t!]
\vspace{3.0in}
\includegraphics{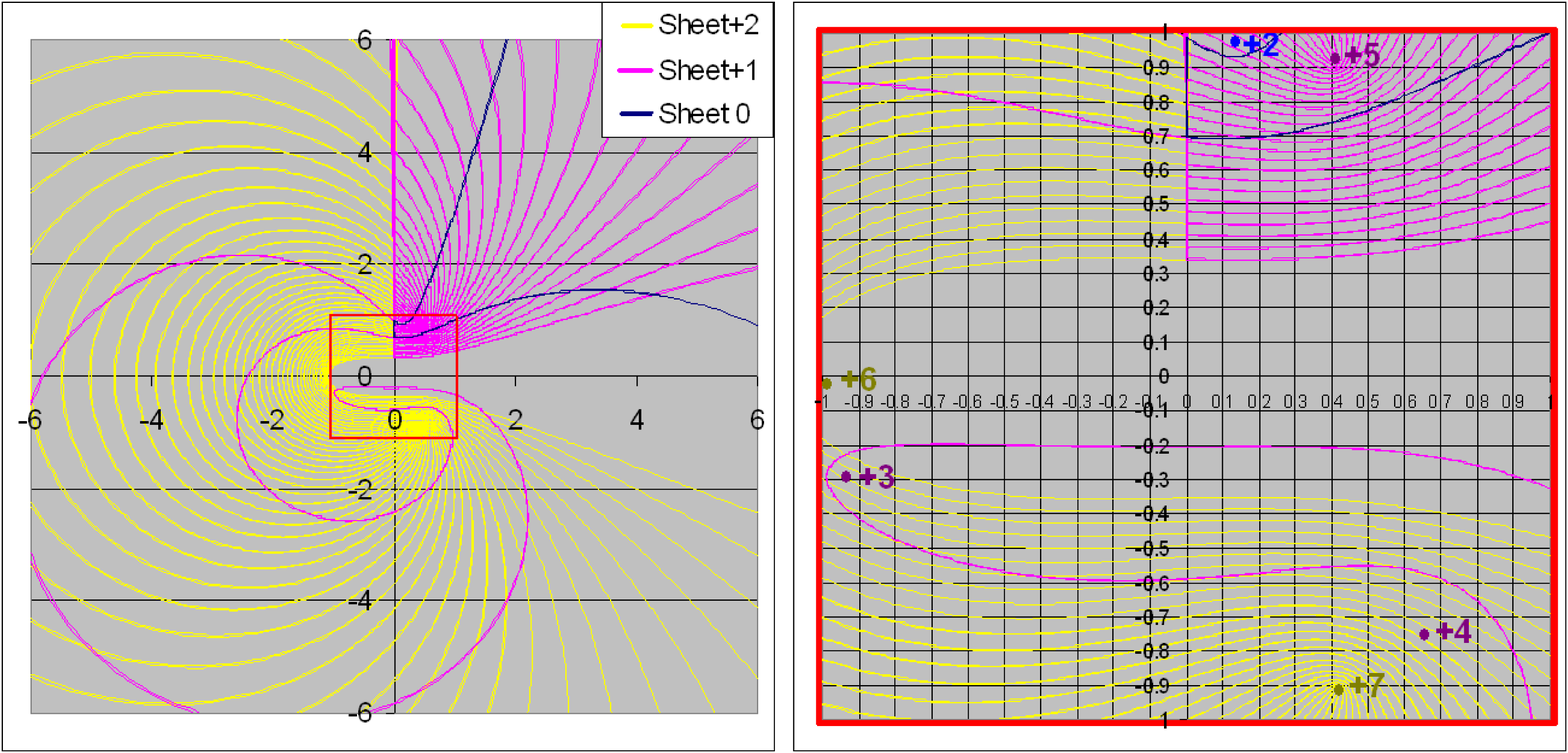}
\caption{The classical orbit in the complex-$x$ plane corresponding to
Fig.~\ref{f10}. The oscillation in Fig.~\ref{f10} corresponds to nested
U-turns around a turning point in the complex plane.}
\label{f11}
\end{figure*}

A characteristic feature of the long orbits is the persistent oscillation in the
classical path which makes huge numbers of U-turns in portions of the complex
plane. These U-turns focus about one of the many complex turning points and
illustrate in a rather dramatic fashion the complex nature of the classical
turning point. (The behavior of real trajectories is much simpler. When a real
trajectory encounters a turning point on the real axis it merely stops and
reverses direction.) In Fig.~\ref{f10} we plot the complex argument of $x(t)$ as
a function of time $t$ for $\epsilon=\pi-2$ and initial condition $x(0)=-17i$.
This orbit has period $T=452.6$ and visits 5 sheets of the Riemann surface. We
show the U-turns of this orbit near a turning point in Fig.~\ref{f11}.

Figures \ref{f10} and \ref{f11} provide a heuristic explanation of how very
long-period orbits arise. In order for a classical trajectory to travel a
great distance in the complex plane, its path must weave through a mine
field of turning points. If the trajectory comes under the influence of a
distant turning point, it executes a huge number of nested U-turns and is
eventually flung back towards its starting point. However, if the initial
condition is chosen very carefully, the complex trajectory can slip past many
turning points before it eventually encounters a turning point that takes
control of the particle. We speculate that it may be possible to find a
special critical
initial condition for which the classical path manages to avoid and slip
past all turning points. Such a path would have an infinitely long period.

\section{Classification of classical orbits}
\label{s3}

In the previous section we explored for a fixed value of $\epsilon$ the
dependence of the classical trajectories on the initial condition. By
varying the initial condition (on the negative imaginary axis) we were able
to produce orbits of incredible topological complexity and with extremely
long periods. In this section we propose a technique for classifying these
orbits. This technique relies on the observation that to calculate the period of
an orbit we may use Cauchy's integral theorem to deform and shrink the orbit
into a curve that tightly encloses the square-root branch cuts that connect the
$\cP\cT$-symmetric pairs of turning points labeled by $n$.

We will argue that all classical orbits having the same period fall into
well-defined topological classes. For example, all three orbits in Fig.~\ref{f2}
have the same period. It is only necessary to examine the {\it central\/} orbits
that terminate at turning points because all other orbits in the same
topological class can be shrunk down to these much simpler central orbits
without changing the period. This simplification allows us to classify all
possible orbits merely by giving the pair of turning points at which the central
orbit terminates.

\begin{figure}[th]\vspace{2.5in}
\includegraphics{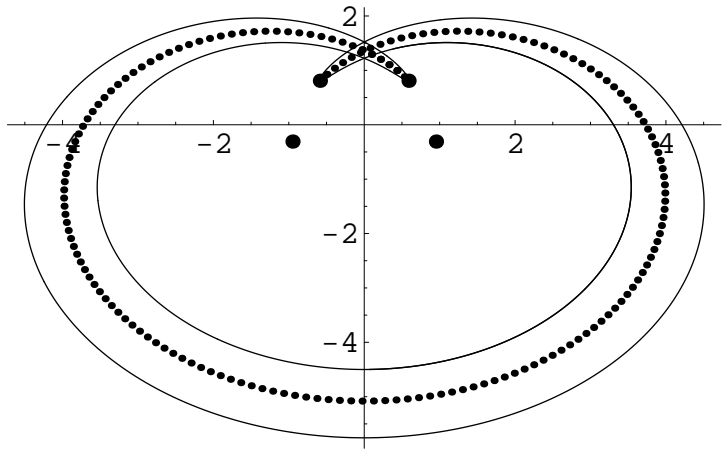}
\caption{Two orbits for $\epsilon=0.5$ that have the same period. The
solid-line orbit encircles the turning points. The dashed-line orbit is the
central orbit that terminates at the $n=1$ pair of turning points. The
$n=0$ and $n=1$ pairs of turning points are indicated by dots. The period of
this class of orbits is $T=5.54559$.}
\label{f12}
\end{figure}

The topological class of orbits shown in Fig.~\ref{f2} is characterized by the
central orbit connecting the $n=0$ pair of turning points. In Fig.~\ref{f12}
we display two classical orbits associated with the $n=1$ pair of turning
points for the case $\epsilon=0.5$. In this figure we show an orbit (solid
line) that encircles the turning points and a central orbit (dashed line),
having the same period, that connects these turning points.

The period of the class of orbits shown in Fig.~\ref{f12} is $T=5.54559$.
To calculate this number we deform the central orbit to a pair of rays that
run from one turning point to the origin and then from the origin to the
other turning point. However, since the turning points lie on different
sheets of the Riemann surface, there are additional contributions from all
other pairs of enclosed turning points. In this case the only other pair of
enclosed turning points is the $n=0$ pair.

In general, there are contributions to the period integral from many enclosed
pairs of turning points. We label each such pair by the integer $j$. The general
formula for the period of a given topological class of classical orbits whose
central orbit terminates on the $n$th pair of turning points is
\begin{eqnarray}
T_n(\epsilon)=2\sqrt{\pi}{\Gamma\left({3+\epsilon\over2+
\epsilon}\right)\over\Gamma\left({4+\epsilon\over4+2\epsilon}\right)}
\sum_{j=0}^{\infty}a_j(n,\epsilon)
\left|\cos\left({(2j+1)\epsilon\pi\over4+2\epsilon}\right)\right|.
\label{e6}
\end{eqnarray}
In this formula the cosines originate from the angular positions of the
turning points in (\ref{e4}). The coefficients $a_j(n,\epsilon)$ are all
nonnegative integers. The $j$th coefficient is nonzero only if the classical
path encloses the $j$th pair of turning points. Each coefficient is an
{\it even\/} integer except for the $j=n$ coefficient, which is an odd
integer. The coefficients $a_j(n,\epsilon)$ satisfy the sum rule
\begin{eqnarray}
\sum_{j=0}^{\infty}a_j(n,\epsilon)=K,
\label{e7}
\end{eqnarray}
where $K$ is the number of times that the central classical path crosses
the imaginary axis. This sum rule truncates the summation in (\ref{e6}) so
that it is only a finite sum. For example, the dashed line in Fig.~\ref{f12}
crosses the imaginary axis three times, so that $K=3$. The formula for the
period of this class of orbits has $a_0=2$ and $a_1=1$.

\begin{figure}[th]\vspace{2.9in}
\includegraphics{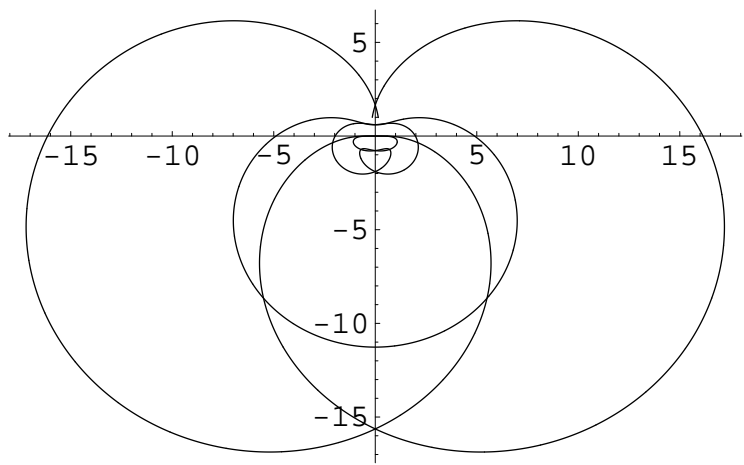}
\caption{Central orbit for $\epsilon=1.149739$ terminating at the $n=1$
pair of turning points. This orbit crosses the imaginary axis 13 times.}
\label{f13}
\end{figure}

As we increase $\epsilon$, the topology of the classical orbits becomes more
complicated. For example, when $\epsilon=1.149739$ the central orbit
belonging to the $n=1$ pair of turning points crosses the imaginary axis 13
times ($K=13$). This orbit is shown in Fig.~\ref{f13}. For this class of
orbits $a_0=2$, $a_1=1$, $a_2=6$, and $a_3=4$. The sum of these coefficients
is 13.

\begin{figure}[th]\vspace{2.6in}
\includegraphics{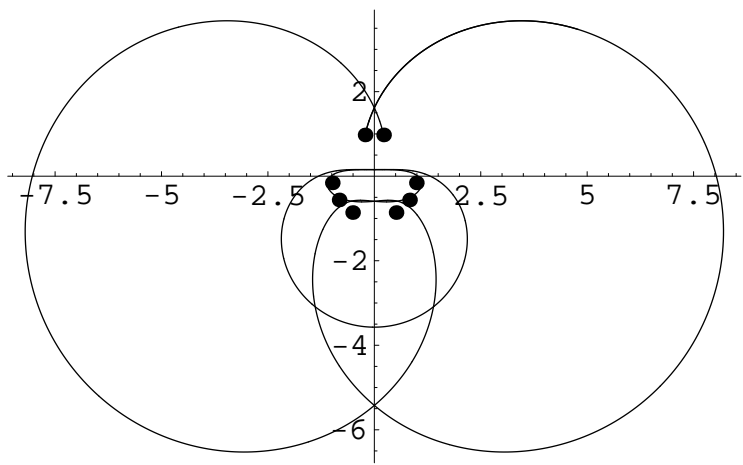}
\caption{Central orbit for $\epsilon=1.225$ terminating at the $n=1$
pair of turning points. This orbit crosses the imaginary axis 9 times.}
\label{f14}
\end{figure}

If we increase $\epsilon$ to $1.225$, the crossing number decreases to
$K=9$. For this class of orbits $a_0=2$, $a_1=1$, $a_2=4$, and $a_3=2$.
The central orbit for this class is shown in Fig.~\ref{f14}.

\begin{figure}[th]\vspace{2.5in}
\includegraphics{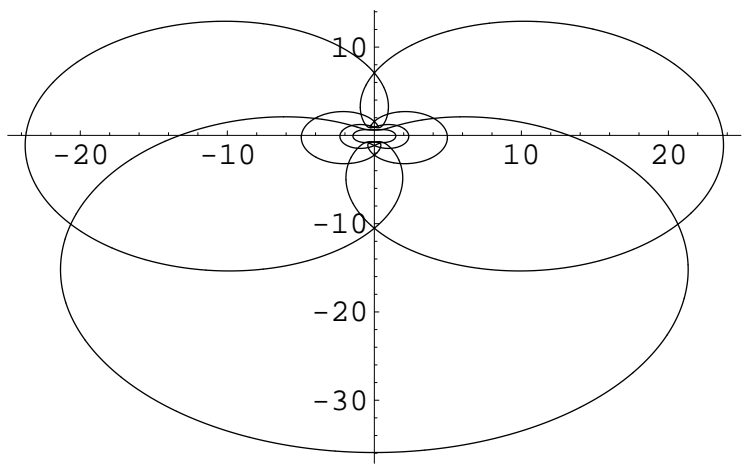}
\caption{Central orbit for $\epsilon=1.3$ terminating at the $n=1$
pair of turning points. This orbit crosses the imaginary axis 17 times.}
\label{f15}
\end{figure}

If we increase $\epsilon$ still further to $1.3$, the crossing number
increases to $K=17$. For this orbit
$a_0=2$, $a_1=3$, $a_2=8$, and $a_3=4$. This orbit is shown in Fig.~\ref{f15}.

\begin{figure}[th]\vspace{2.4in}
\includegraphics{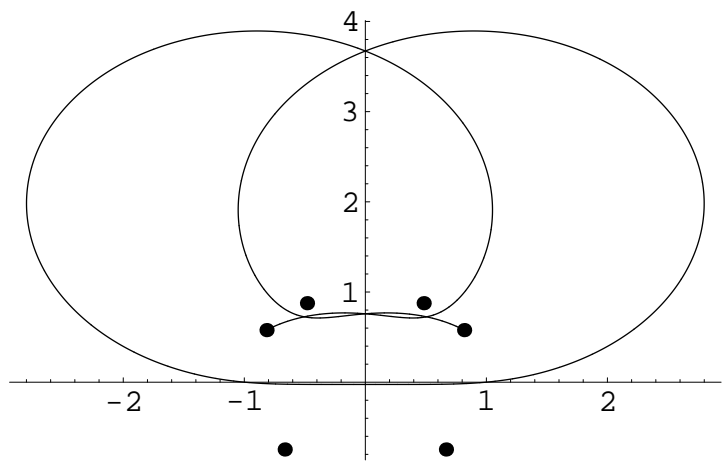}
\caption{Central orbit for $\epsilon=2.31$ terminating at the $n=1$
pair of turning points. This orbit crosses the imaginary axis 5 times.
Note that this orbit does not enclose the $n=0$ turning points.}
\label{f16}
\end{figure}

For $\epsilon=2.31$ the number of crossings decreases again to $K=5$.
For this orbit $a_0=0$, $a_1=3$, and $a_2=2$. This orbit is shown in
Fig.~\ref{f16}. Note that unlike the orbits in Figs.~\ref{f15} and \ref{f14}
this orbit does not enclose the $n=0$ turning points. This is why the 
$a_0$ coefficient vanishes.

\begin{figure}[th]\vspace{2.5in}
\includegraphics{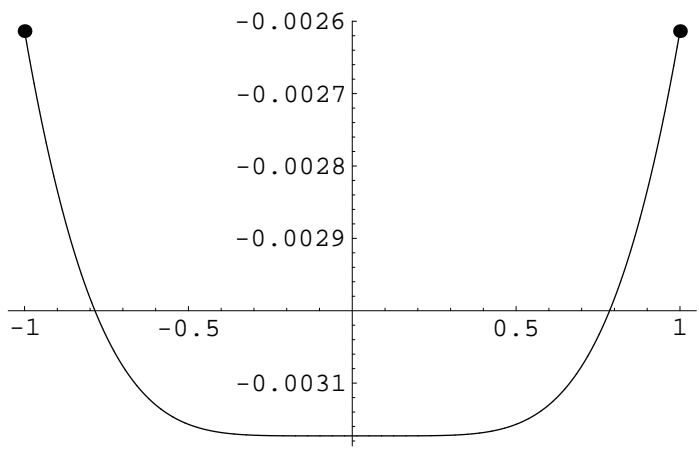}
\caption{Central orbit for $\epsilon=4.01$ terminating at the $n=1$
pair of turning points. This orbit crosses the imaginary axis only once.}
\label{f17}
\end{figure}

If we continue to increase the value of $\epsilon$, the topology of the
classical orbits eventually simplifies. For all $\epsilon\geq4$ we find that
$K=1$. For example, in Fig.~\ref{f17} we illustrate the central orbit for
$\epsilon=4.01$. For this class of orbits we have $a_1=1$ and all other
coefficients vanish.

\begin{figure}[th]\vspace{1.9in}
\includegraphics{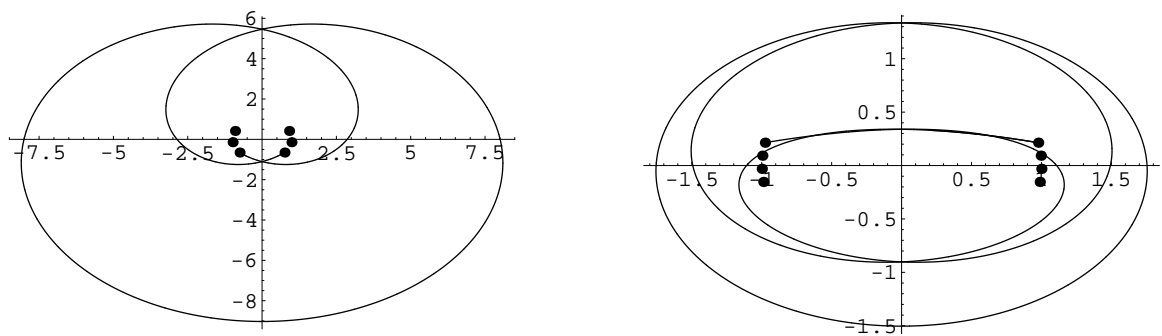}
\caption{On the left the central orbit for $\epsilon=0.2$ terminating at the
$n=2$ pair of turning points. This orbit crosses the imaginary axis 5 times.
On the right the central orbit for $\epsilon=0.04$ terminating at the
$n=3$ pair of turning points. This orbit crosses the imaginary axis 7 times.}
\label{f18a}
\end{figure}

For small $\epsilon$ the classical orbits terminating at the $n=2$ and $n=3$
turning points behave in a similar fashion. When $\epsilon=0.2$, the $n=2$
central
orbit crosses the imaginary axis 5 times and when $\epsilon=0.04$, the $n=3$
central orbit crosses the imaginary
axis 7 times (see Fig.~\ref{f18a}). In the former case $a_0=2$, $a_1=2$, and
$a_2=1$ and in the latter case $a_0=2$, $a_1=2$, $a_2=2$, and $a_3=1$.

\begin{figure*}[th]\vspace{0.0in}
\includegraphics{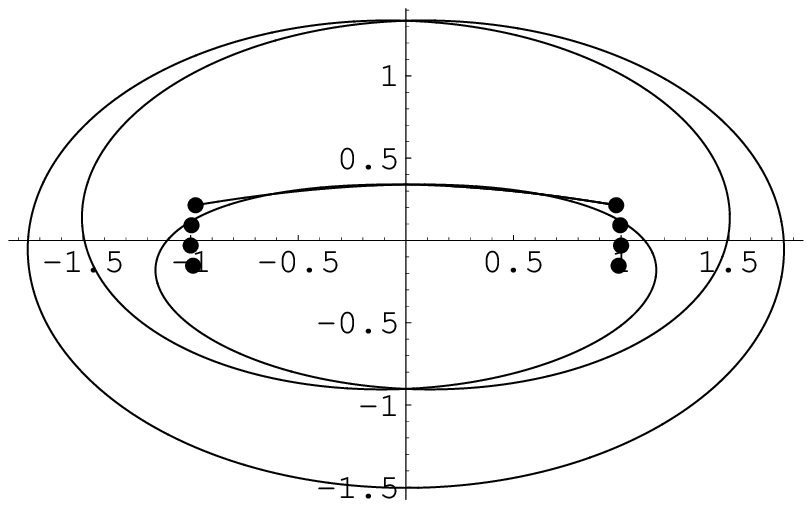}
\end{figure*}

\begin{figure}[th]\vspace{2.5in}
\includegraphics{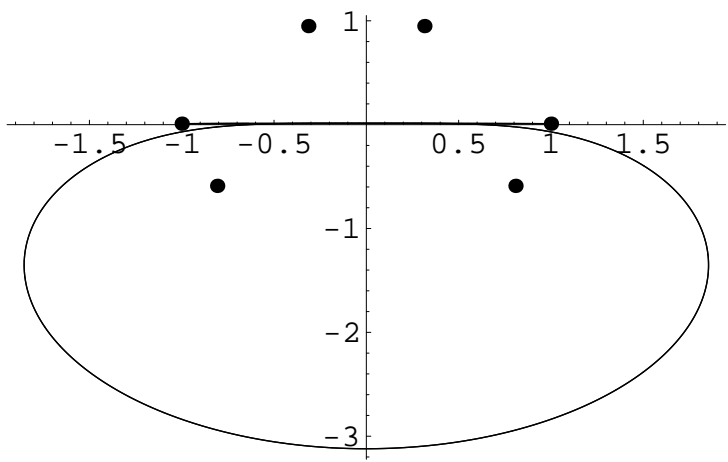}
\caption{Central orbit for $\epsilon=1.34$ terminating at the $n=2$
pair of turning points. This orbit crosses the imaginary axis 3 times.
This orbit does not enclose the $n=1$ turning points.}
\label{f19}
\end{figure}

As $\epsilon$ increases, the topology in Fig.~\ref{f18a} changes. For
example, when $\epsilon=1.34$, the $n=2$ central orbit crosses the
imaginary axis 3 times and we find that $a_0=2$, $a_1=0$, and $a_2=1$
(see Fig.~\ref{f19}).

We can see from Figs.~\ref{f12} - \ref{f19} that a clear pattern emerges. When
$\epsilon$ is small (less than $\frac{1}{n}$) the central path that joins the
$n$th pair of turning points crosses the imaginary axis $K=2n+1$ times and the
coefficients
$a_j$ have a simple pattern: $a_j=2$ for $j<n$ and $a_n=1$. When $\epsilon$ is
large, specifically $\epsilon>4n$, the topology of the central classical
orbits becomes extremely simple and there is only one crossing ($K=1$). For this
case $a_n=1$. The most interesting behavior occurs for intermediate values of
$\epsilon$, where we observe remarkable transitions as a function of $\epsilon$
that exhibit critical behavior. This behavior is discussed in Sec.~\ref{s4}.

\section{Critical behavior in $\epsilon$}
\label{s4}
In this section we study the behavior of the classical trajectories as the
parameter $\epsilon$ is varied. We restrict our attention to the central orbits
(the closed orbits that terminate at turning points).

\begin{figure}[th]\vspace{1.9in}
\includegraphics{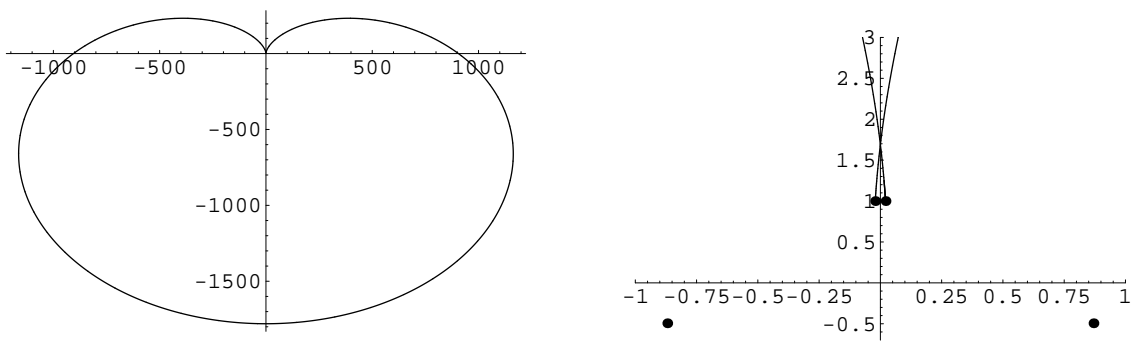}
\caption{On the left the central orbit for $\epsilon=0.98$ terminating at the
$n=1$ pair of turning points. This orbit crosses the imaginary axis 3 times.
Note that this orbit is huge and that its shape resembles a cardioid. Our
numerical studies indicate that as $\epsilon$ tends to 1, the size of the orbit
becomes infinite. On the right is an enlargement showing the detail near the
origin.}
\label{f20a}
\end{figure}

We begin by considering the case of central orbits that terminate at the
$n=1$ pair of turning ponts. For $\epsilon<1$ all orbits cross the imaginary
axis three times ($K=3$). However, as $\epsilon$ approaches 1 from below, the
orbits become huge and cardioid shaped, as illustrated in Fig.~\ref{f20a}. The
vertical and horizontal extent of this orbit is about $2,000$.

\begin{figure*}[th]\vspace{0.0in}
\includegraphics{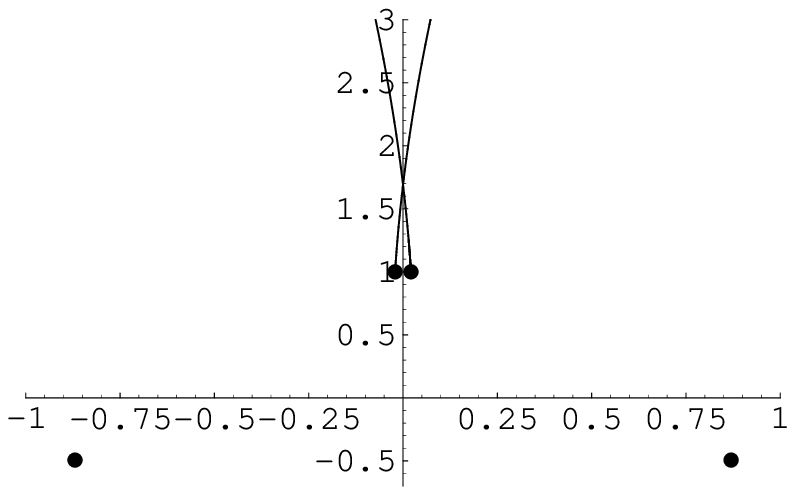}
\end{figure*}

We have done an extremely detailed numerical study to determine the precise
value of $\epsilon$ at which the size of the orbit becomes infinite. In
particular, we have determined the point $y(\epsilon)$ at which the orbit
crosses the negative imaginary axis. [The number $y(\epsilon)$ becomes large and
negative.] We find that a very good fit to $y(\epsilon)$ when $\epsilon$ is just
below 1 is given by
\begin{equation}
y(\epsilon)=-a(b-\epsilon)^{-\gamma}.
\label{e8}
\end{equation}
By fitting this formula to a large set of plots, we find that
$b\approx0.999947$, and we therefore assume that the exact value of $b$ is 1.
The values of the other two parameters are $a\approx0.883032$ and
$\gamma\approx1.93757$. Thus, as $\epsilon$ approaches 1 from below, we
observe critical behavior characterized by the index $\gamma$. (We note that
it is at $\epsilon=1$ that the angular distance between successive turning
points becomes sufficiently small that the $n=1$ pair of turning points
enters the principal sheet of the Riemann surface. However, we do not
understand why this might cause the observed critical behavior.)

\begin{figure}[th]\vspace{2.0in}
\includegraphics{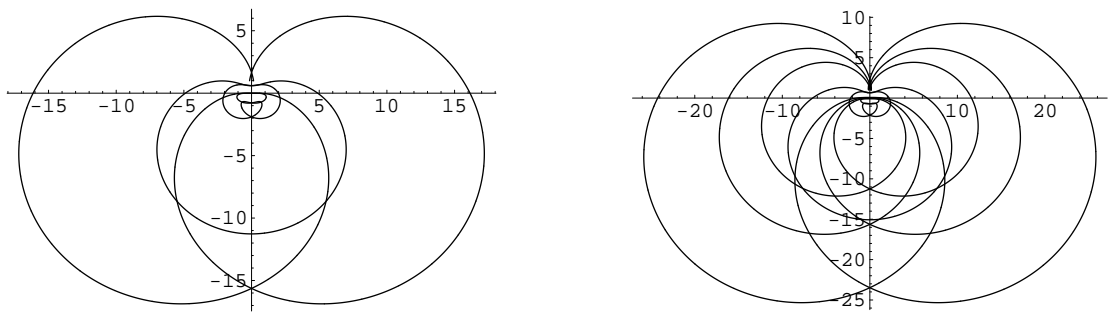}
\caption{Transition between two topological classes of graphs.
The left graph shows the central orbit for $\epsilon=1.149739$ terminating at
the $n=1$ pair of turning points. This orbit crosses the imaginary axis 13
times. The right graph displays the central orbit for $\epsilon=1.149738$
terminating at the $n=1$ pair of turning points. This orbit crosses the
imaginary axis 25 times.}
\label{f21a}
\end{figure}

When $\epsilon$ is slightly larger than 1 it is too difficult to discern the
topological structure of the orbits because these orbits are so large and
complicated that they overwhelm the numerical capability of the computer. 
However, we can construct numerically the orbits for $\epsilon$ larger than
about $1.14736$. We observe a remarkable behavior in the structure of the
orbits as we approach this value of $\epsilon$ from above
in the region $1.14736<\epsilon<1.169$. Specifically, when $\epsilon>1.1497389$,
the $n=1$ central orbits have $K=13$ crossings as illustrated in Fig.~\ref{f21a}
on the left. This figure shows the $K=13$ orbit for $\epsilon=1.149739$.
When $\epsilon$ decreases slightly to the value $\epsilon=1.149738$ we
observe a transition to an orbit with $K=25$ crossings, as shown in
Fig.~\ref{f21a} on the right.

\begin{figure*}[th]\vspace{0.0in}
\includegraphics{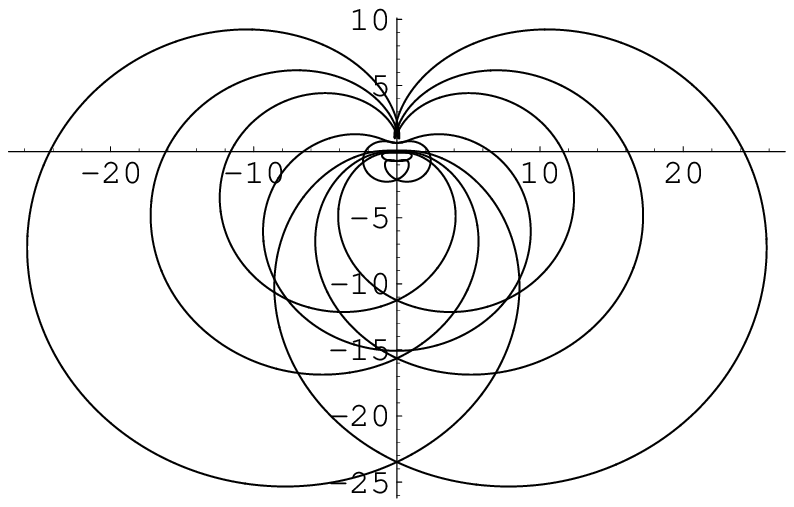}
\end{figure*}

\begin{figure}[th]\vspace{2.0in}
\includegraphics{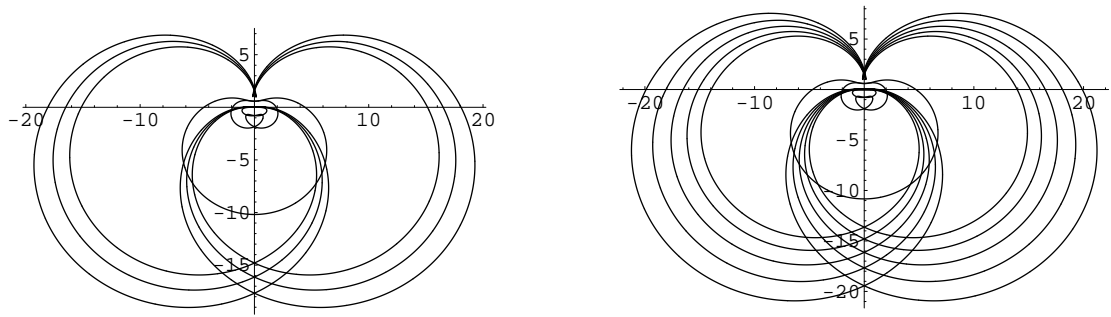}
\caption{Second transition between two topological classes of graphs.
The left graph shows the central orbit for $\epsilon=1.1478263$ terminating at
the $n=1$ pair of turning points. This orbit crosses the imaginary axis 25
times. The right graph shows the central orbit for $\epsilon=1.1478262$
terminating at the $n=1$ pair of turning points.
This orbit crosses the imaginary axis 37 times.}
\label{f22a}
\end{figure}

\begin{figure}[th]\vspace{2.8in}
\includegraphics{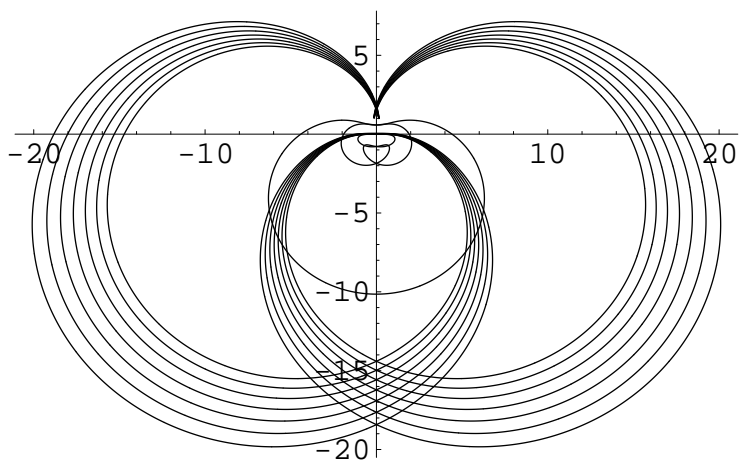}
\caption{Central orbit for $\epsilon=1.1475$ terminating at the $n=1$ pair of
turning points. This orbit crosses the imaginary axis 49 times.}
\label{f23}
\end{figure}

\begin{figure}[th]\vspace{2.6in}
\includegraphics{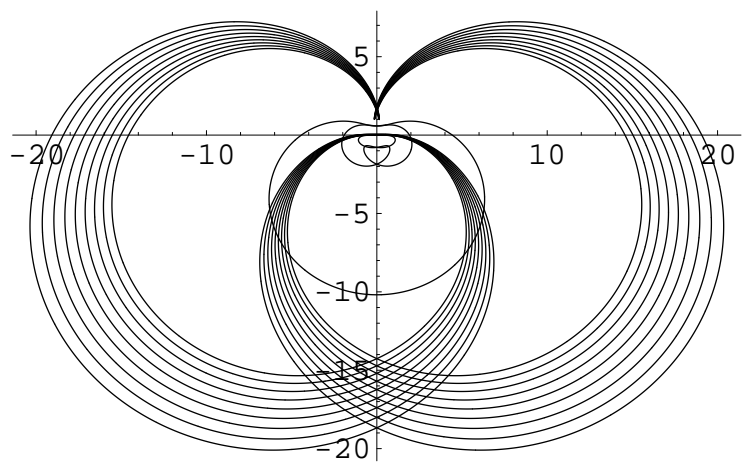}
\caption{Central orbit for $\epsilon=1.14745$ terminating at the $n=1$ pair
of turning points. This orbit crosses the imaginary axis 61 times.}
\label{f24}
\end{figure}

\begin{figure}[th]\vspace{2.8in}
\includegraphics{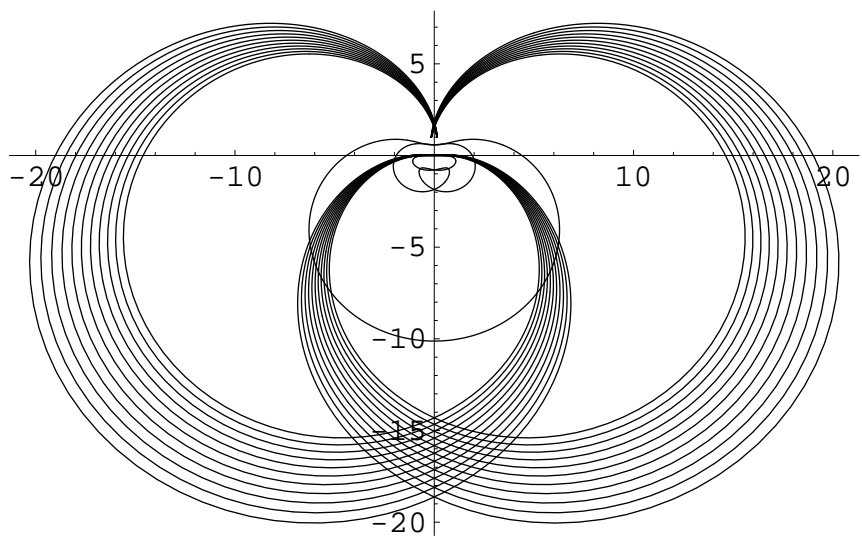}
\caption{Central orbit for $\epsilon=1.1474$ terminating at the $n=1$ pair of
turning points. This orbit crosses the imaginary axis 73 times.}
\label{f25}
\end{figure}

\begin{figure}[th]\vspace{2.9in}
\includegraphics{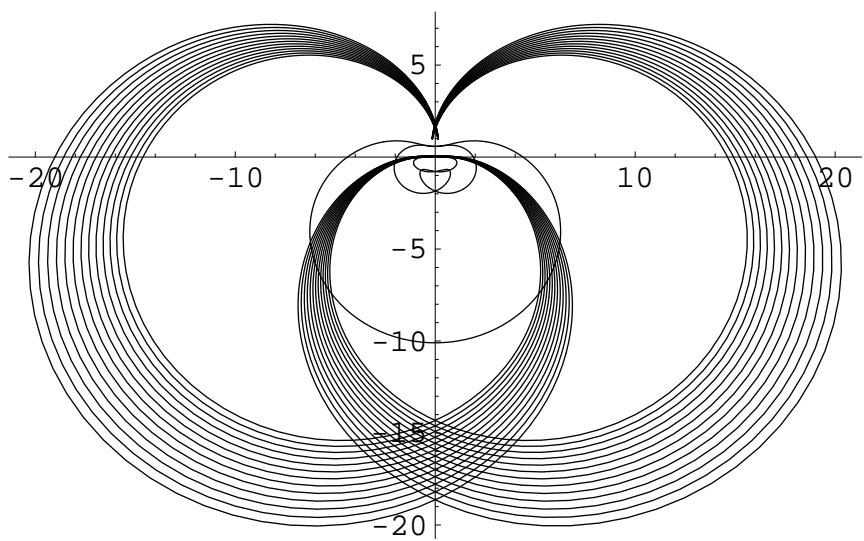}
\caption{Central orbit for $\epsilon=1.14737$ terminating at the $n=1$ pair of
turning points. This orbit crosses the imaginary axis 85 times.}
\label{f26}
\end{figure}

As we continue to decrease $\epsilon$, we encounter a second transition from
orbits having $K=25$ crossings to orbits having $K=37$ crossings. This
transition occurs very near the value $\epsilon=1.14782625$. To illustrate this
transition we show in Fig.~\ref{f22a} the $n=1$ orbits for
$\epsilon=1.1478263$ (left) and for $\epsilon=1.1478262$ (right).
\begin{figure*}[th]\vspace{0.0in}
\includegraphics{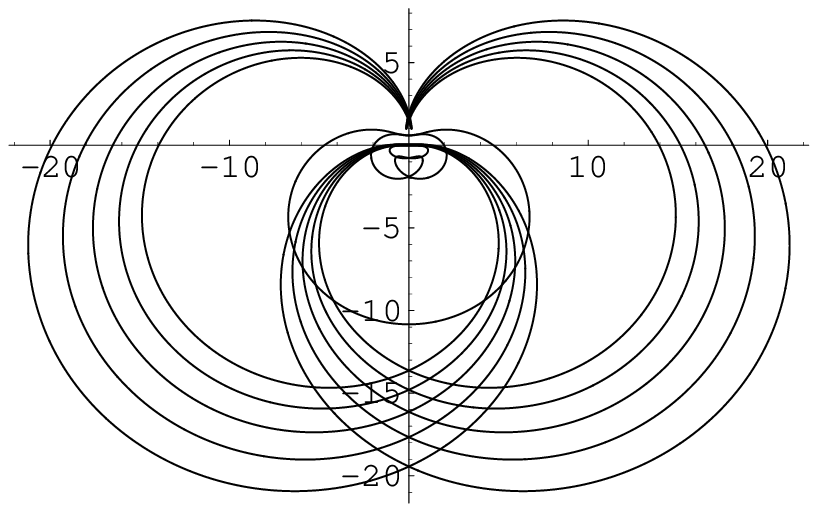}
\end{figure*}
As $\epsilon$ continues to decrease, there is a third transition at about
$\epsilon=1.14756475$ where the crossing number of the central orbits
jumps from $K=37$ to $K=49$ (see Fig.~\ref{f23}).
A fourth transition occurs at $\epsilon=1.14746085$ at which the $n=1$
orbits jump from $K=49$ crossings to $K=61$ crossings (see Fig.~\ref{f24}).
We observe a fifth and sixth transition from $K=61$ to $K=73$ (see
Fig.~\ref{f25}) and from $K=73$ to $K=85$ (see Fig.~\ref{f26}). It is clear
that with each transition the value of $K$ increases by 12. The accumulation
point of these transitions is close to $\epsilon=1.14736$. Throughout the region
illustrated by Figs.~\ref{f21a} -\ref{f26} there is an exact closed-form
expression for the coefficients $a_j$ appearing in the formula (\ref{e6}) for
the period $T$. In general, if we express the crossing number $K$ in the form
$K=13+12k$ ($k=0,~1,~2,~\dots$), then
\begin{equation}
(a_0,a_1,a_2,a_3,a_4,a_5)=(2,1+2k,6+4k,4+4k,0,2k)
\label{e9}
\end{equation}
with all higher coefficients vanishing.

The region of $\epsilon$ between $\epsilon=1$ and $\epsilon=1.14736$ remains
largely unexplored and its behavior is mysterious. We have been able to find
isolated values of $\epsilon$ in this region for which we can determine the
orbit numerically. (One such value is $\epsilon=1.03$, and for this value we
find that $K=29$.) However, we do not understand how the topology of the orbits
depends on $\epsilon$ in this region.

What happens to the $K=13$ orbits when $\epsilon$ increases? From a detailed
numerical analysis of the classical orbits, we find a new
critical point near $\epsilon=1.16898$. As $\epsilon$ approaches this
critical point from below, the orbits on the principal sheet of the Riemann
surface again resemble huge cardioids and the intercept on the negative
imaginary axis is given by (\ref{e8}) with $a=0.013159$, $b=1.16898$, and
the index $\gamma=1.70439$.

\begin{figure}[th]\vspace{2.0in}
\includegraphics{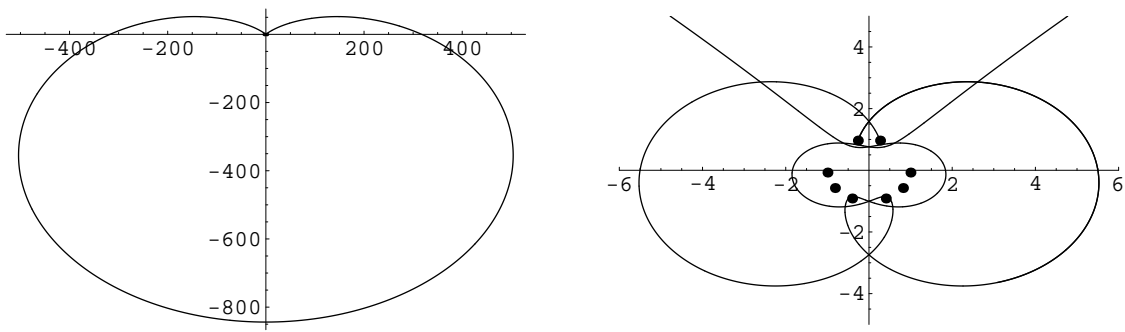}
\caption{Central orbit for $\epsilon=1.282$ terminating at the $n=1$
pair of turning points. This orbit crosses the imaginary axis 9 times.
To the right is a detail of the region near the origin.}
\label{f27a}
\end{figure}

Beyond this value of $\epsilon$ we encounter a new and unexplored mysterious
region. The upper boundary of this region is near $\epsilon=1.21$. As we
approach this value of $\epsilon$ from above, we again observe a sequence of
transitions in which the crossing number $K$ again appears to jump
arithmetically. One orbit in this region is shown in Fig.~\ref{f27a} and
another is shown in Fig.~\ref{f14}. These orbits cross the imaginary axis $K=9$
times.

\begin{figure*}[th]\vspace{0.0in}
\includegraphics{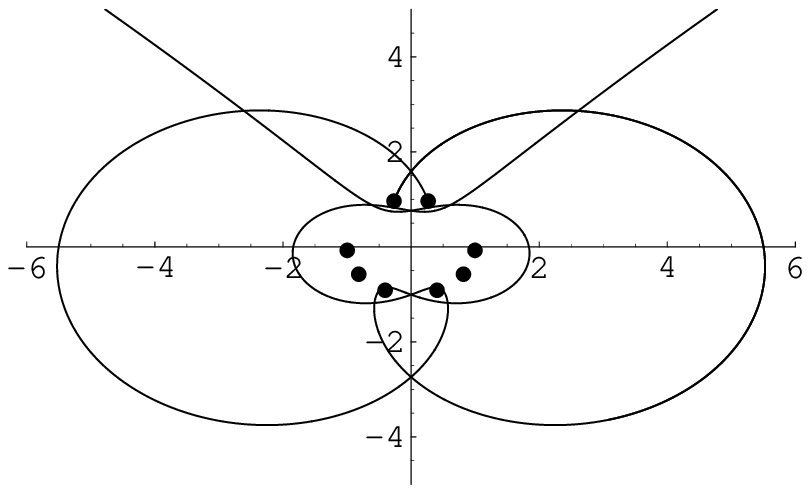}
\end{figure*}

\begin{figure}[th]\vspace{1.9in}
\includegraphics{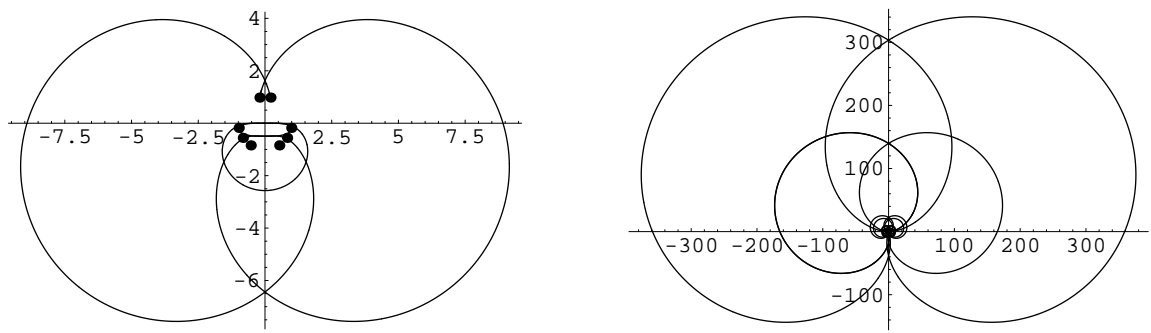}
\caption{Transition between a $K=9$ orbit (left) and a $K=69$ orbit (right. The
left graph shows a central $n=1$ orbit that corresponds to $\epsilon=1.2115215$
and the right graph shows a central $n=1$ orbit that corresponds to
$\epsilon=1.2115214$. It is not easy to see that the right orbit actually
crosses the imaginary axis 69 times, so the region of this orbit near the
origin is displayed in Fig.~\ref{f29a}.}
\label{f28a}
\end{figure}

\begin{figure}[th]\vspace{1.9in}
\includegraphics{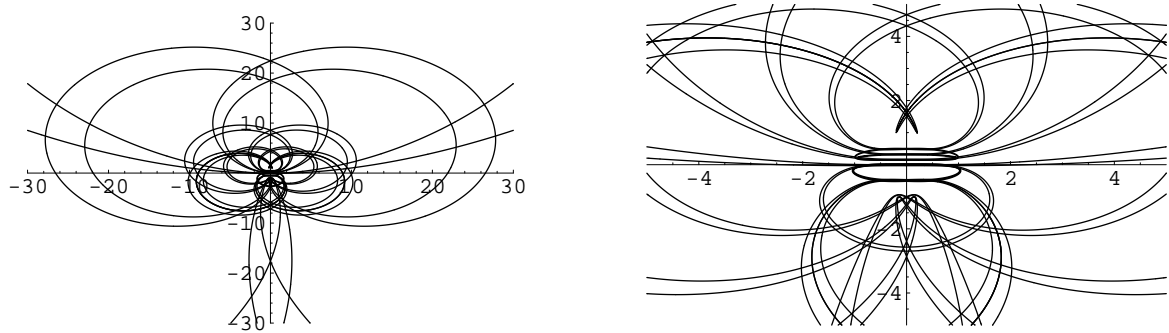}
\caption{Two enlargements of the $n=1$ central orbit for
$\epsilon=1.2115214$ in Fig.~\ref{f28a} (right). By careful examination of
these detailed graphs, one can determine that the crossing number is
$K=69$.}
\label{f29a}
\end{figure}

The first transition in this region of classical orbits occurs very near
$\epsilon=1.21152145$. At this transition the value of the crossing number
jumps from $K=9$ to $K=69$, and we observe that this change is a multiple of 12.
Classical orbits just above and below this transition are shown in
Fig.~\ref{f28a}.
\begin{figure*}[th]\vspace{0.0in}
\includegraphics{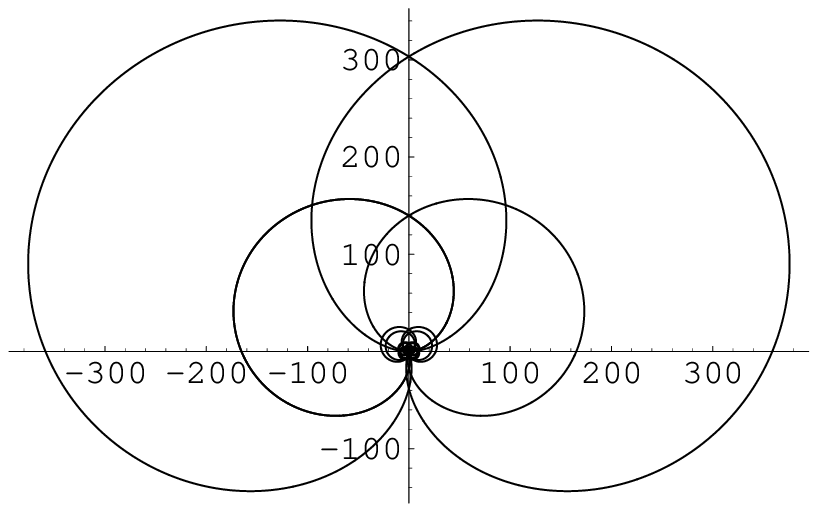}
\end{figure*}
It is very difficult to see that the orbit on the right in Fig.~\ref{f28a}
actually crosses the imaginary axis 69 times, so in Fig.~\ref{f29a} we show
two enlargements of the region near the origin.

\begin{figure*}[th]\vspace{0.0in}
\includegraphics{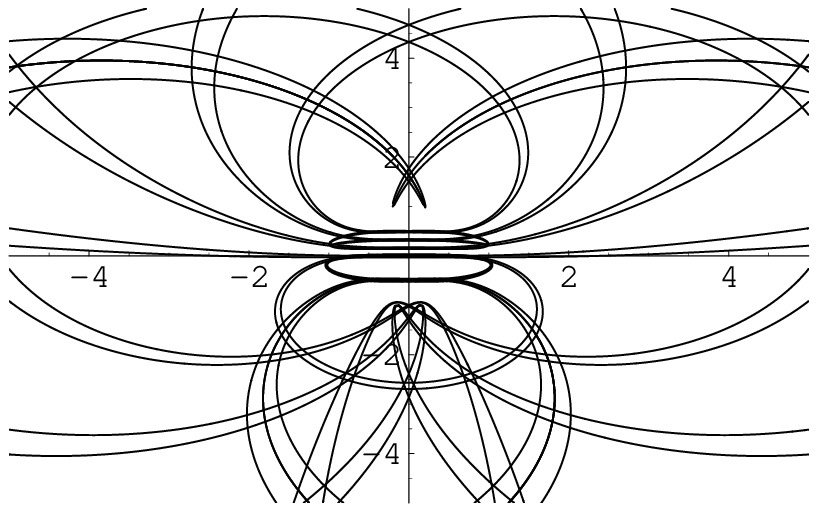}
\end{figure*}

As we continue to increase $\epsilon$, we encounter a third critical
transition that marks the upper end of this region. At this transition the
classical orbits on the principal sheet again resemble huge cardioids and
again the intercept on the imaginary axis is well approximated by the
formula (\ref{e8}) with $a=0.0423448$, $b=1.2837$, and the critical index
$\gamma=1.55255$.

\begin{figure}[th]\vspace{1.9in}
\includegraphics{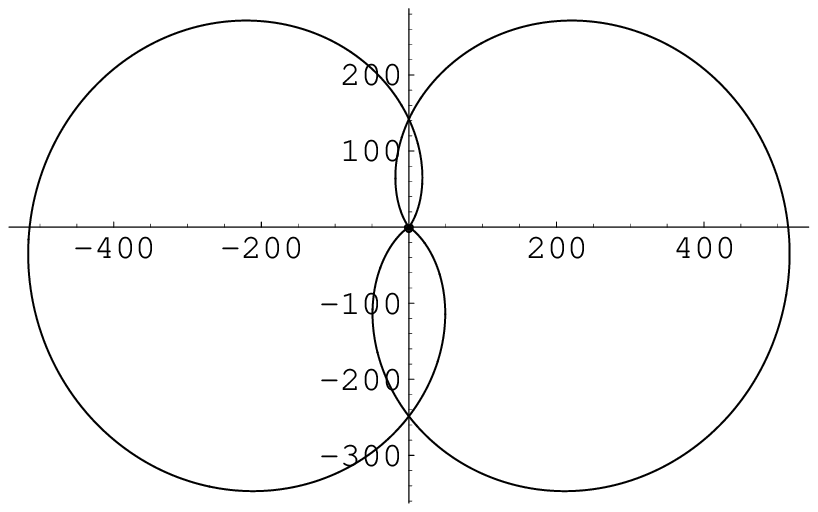}
\caption{Central orbit for $\epsilon=1.286$ terminating at the $n=1$
pair of turning points. This orbit crosses the imaginary axis 17 times.
The central orbit is dominated by a huge double-cardioid structure having a
horizontal extent of over 1000. This structure lies on the $\pm1$ sheets of
the Riemann surface. To the right is a detail of the region near the origin.}
\label{f30a}
\end{figure}

\begin{figure}[th]\vspace{1.8in}
\includegraphics{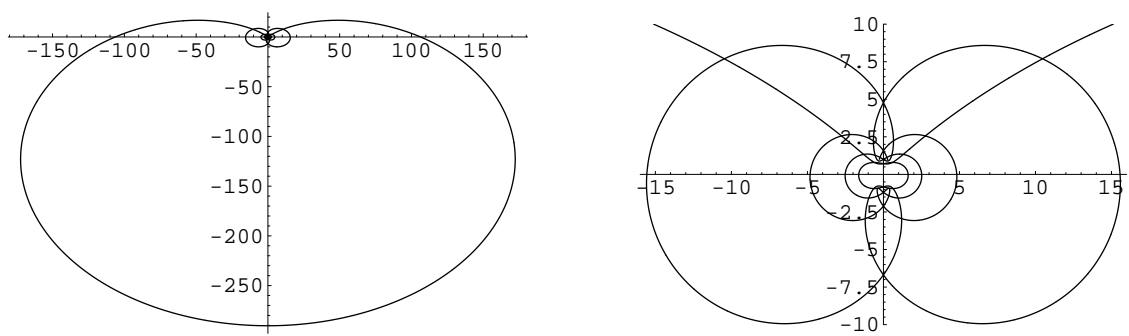}
\caption{Central orbit for $\epsilon=1.305$ terminating at the $n=1$
pair of turning points. This orbit crosses the imaginary axis 17 times.
On the principal sheet the central orbit resembles a large cardioid having a
horizontal extent of over 300.
To the right is a detail of the region near the origin.}
\label{f31a}
\end{figure}

Surprisingly, we find a new kind of critical behavior above this region.
For $\epsilon$ between the values $1.284$ and $1.306$ we find $n=1$
classical orbits with crossing number $K=17$. We have already shown one
such orbit Fig.~\ref{f15}. As $\epsilon$ approaches the lower boundary of this
region, the classical orbits become
huge, but do not resemble cardioids. Rather, they resemble {\it double\/}
cardioids -- that is, cardioids having two notches instead of one. An example of
a $K=17$ orbit near the lower edge of this region is shown in Fig.~\ref{f30a}.
As $\epsilon$ approaches the upper boundary of this region, the classical orbits
once again resemble huge cardioids. An example of a $K=17$ orbit near the upper
edge of this region is shown in Fig.~\ref{f31a}.

\begin{figure*}[th]\vspace{0.0in}
\includegraphics{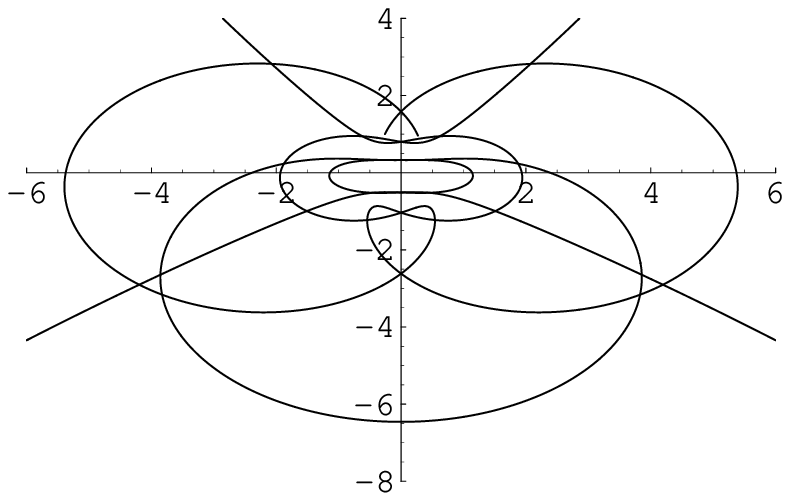}
\end{figure*}

\begin{figure*}[th]\vspace{0.0in}
\includegraphics{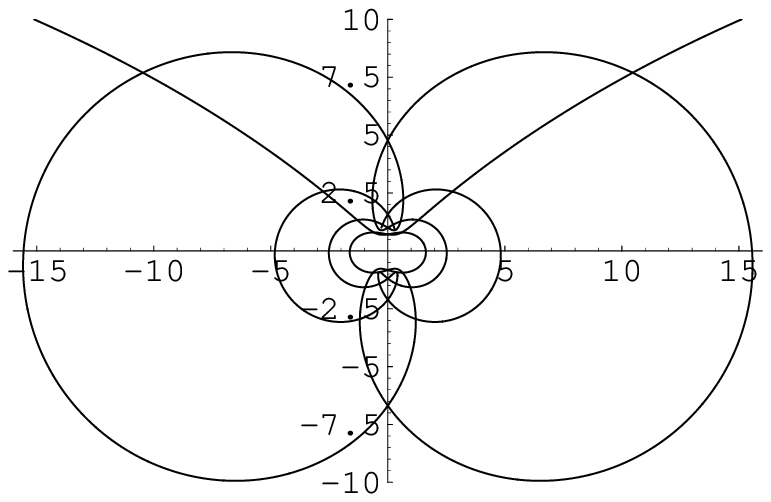}
\end{figure*}

The same sort of behavior characterized by narrow regions bounded by critical 
points is observed for orbits terminating on $n=2$ and higher critical points. 
Numerical studies indicate that the first critical point for $n=2$ is near
$\epsilon=\frac{1}{2}$ and the first critical point for $n=3$ is approximately
at $\epsilon=\frac{1}{3}$. As $\epsilon$ approaches these critical points from 
below, we observe the same kind of critical behavior that is expressed in
(\ref{e8}). For the case $n=2$, we find that $a=0.071388$, $b=0.499759$, and the
critical index $\gamma=3.7063$ and for the case $n=3$, we find that $a=
0.008258$, $b=0.3341$, and the critical index $\gamma=5.17017$. While it is
difficult to assess the precise numerical accuracy of these results, we believe
that it is safe to conjecture that as a function of $n$, $b=\frac{1}{n}$.
Furthermore, $a$ appears to decay geometrically with increasing $n$ and $\gamma$
seems to grow arithmetically with increasing $n$.

\begin{figure}[th]\vspace{2.0in}
\includegraphics{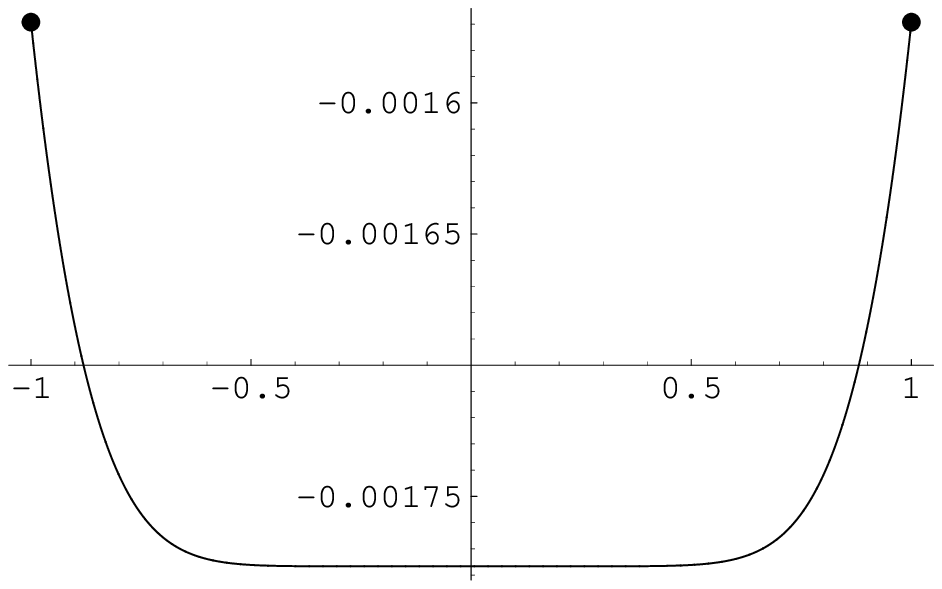}
\caption{On the left, central orbit terminating at the $n=2$ pair of turning
points for $\epsilon=8.01$. On the right, central orbit terminating at the $n=3$
pair of turning points for $\epsilon=12.01$.}
\label{f32a}
\end{figure}

\begin{figure}[th]\vspace{2.0in}
\includegraphics{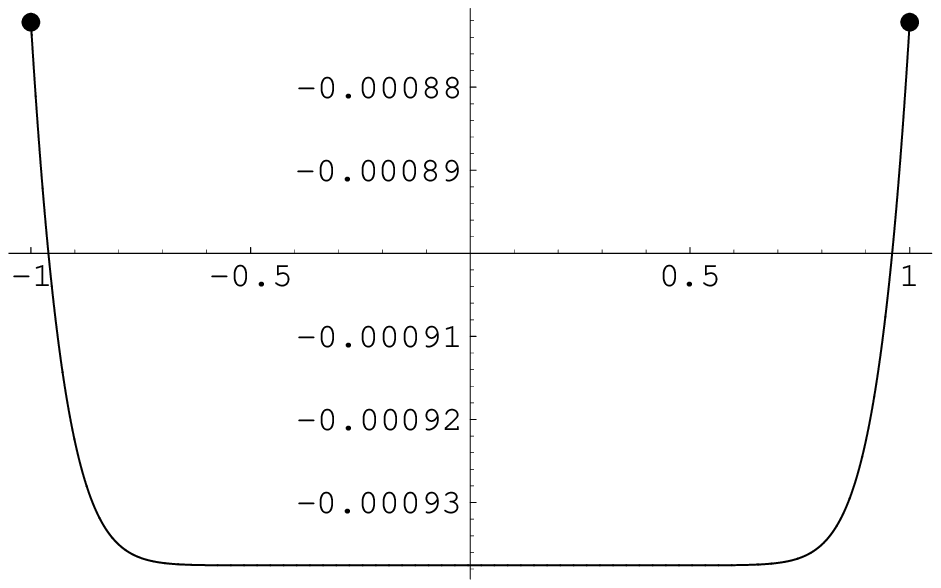}
\caption{On the left, central orbit terminating at the $n=4$ pair of turning
points for $\epsilon=16.01$. On the right, central orbit terminating at the
$n=5$ pair of turning points for $\epsilon=20.01$.}
\label{f33a}
\end{figure}

Eventually, this extraordinarily complicated array of regions in $\epsilon$,
which are bounded by critical points, gives way to a very simple and almost
featureless behavior. We find that when $\epsilon>4n$ the classical
trajectories lie entirely on the principal sheet of the Riemann surface and
cross the imaginary axis exactly once. We have already seen in Fig.~\ref{f17} 
an example of this simple behavior. Four additional examples of these central
classical trajectories for $\epsilon$ lying just above this transition are shown
in Figs.~\ref{f32a} and \ref{f33a}. The transition to simple behavior occurs
just as the $n$th pair of turning points crosses the real axis on the principal
sheet of the Riemann surface.

\begin{figure*}[th]\vspace{0.0in}
\includegraphics{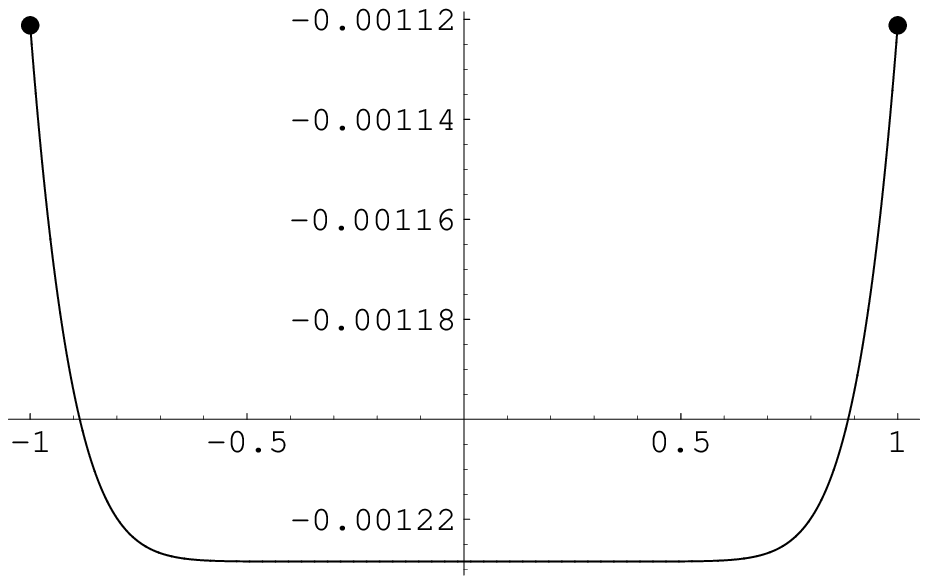}
\end{figure*}

\begin{figure*}[th]\vspace{0.0in}
\includegraphics{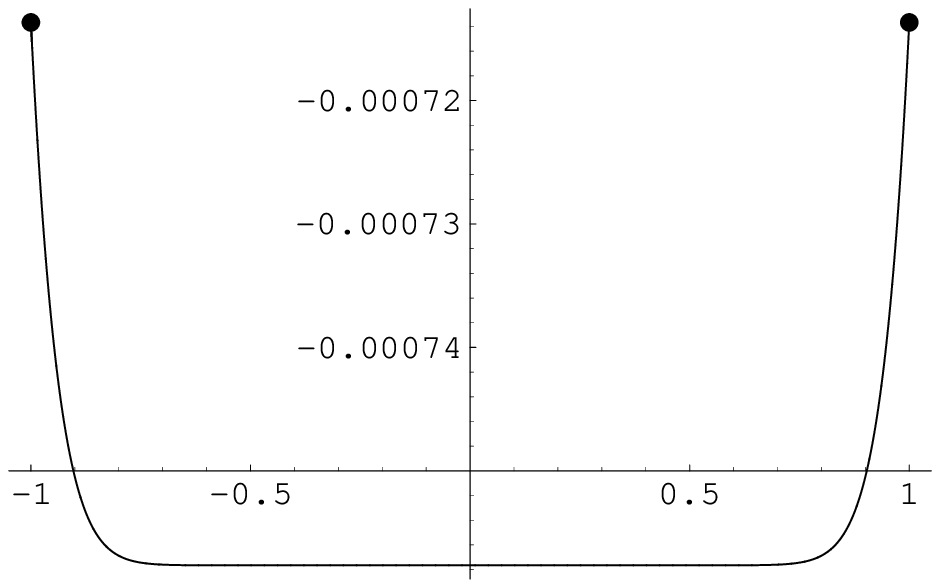}
\end{figure*}

\begin{figure}[th]\vspace{2.0in}
\includegraphics{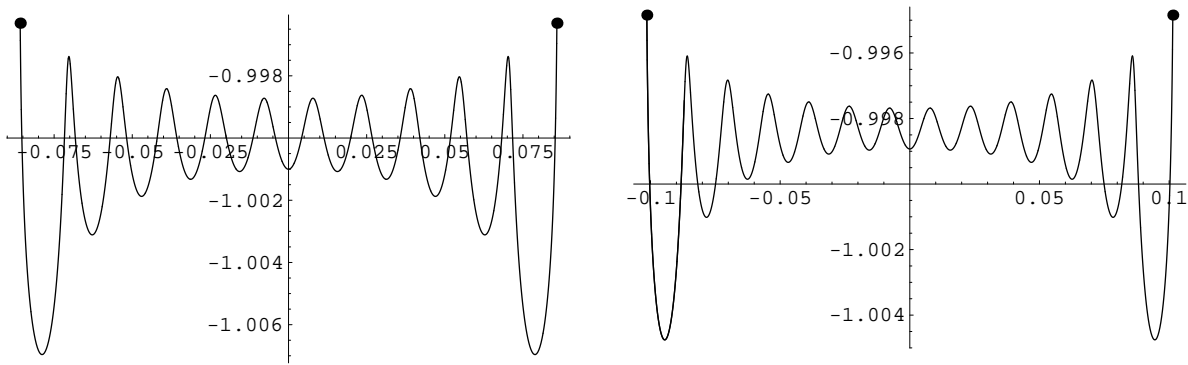}
\caption{On the left, central curve terminating at the $n=5$ pair of turning
points for $\epsilon=400$. On the right, central curve terminating at the $n=6$
pair of turning points for $\epsilon=400$. In general, the number of
oscillations is $2n+1$.}
\label{f34a}
\end{figure}

While the orbits shown in Figs.~\ref{f32a} and \ref{f33a} appear to have no
interesting structure, in fact they exhibit an interesting low-amplitude 
oscillation. To see this oscillation we take $\epsilon$ large and plot the
central curves for two values of $n$ in Fig.~\ref{f34a}. Evidently, the
number of oscillations is $2n+1$.

\begin{figure*}[th]\vspace{0.0in}
\includegraphics{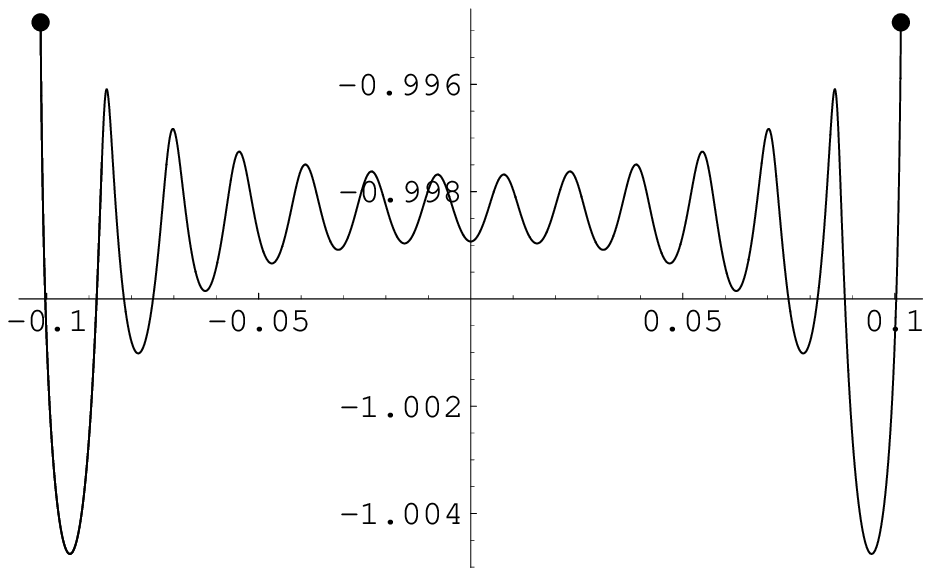}
\end{figure*}

\section{Concluding remarks}
\label{s5}
This paper is a descriptive taxonomy of possible behaviors of classical
trajectories of a particle that obeys the Hamiltonian (\ref{e1}). It is
surprising how rich and elaborate these behaviors can be. We have found
classical paths that are extremely sensitive to changes in initial
conditions. We have also shown that the classical paths are delicately
dependent on the value of $\epsilon$, so highly dependent that they exhibit
critical behavior. We have also identified many problems that require further
investigation, both numerical and analytic. For example, we do not understand
the true nature of the multiple critical behaviors that we have discovered.

The classical behavior that we have found is in part remiscent of the
period-lengthening route to chaos that is observed in logistic maps. In the case
of logistic maps the critical behavior is a function of a multiplicative
parameter $\lambda$, while here the parameter $\epsilon$ appears in the exponent
of a complex function. It may be possible to regard the rich behavior
described in this paper as a kind of complex extension of chaos theory.

\ack
KAM thanks the Physics Department at Washington University for its hospitality.
CMB and KAM are supported by the US Department of Energy.


\section*{References}
\begin{thebibliography}{999}
\bibitem{r1} C.~M.~Bender and S.~Boettcher, Phys.~Rev.~Lett.~{\bf 80}, 5243
(1998).

\bibitem{r2} P.~Dorey, C.~Dunning, and R.~Tateo, J.~Phys.~A:
Math.~Gen.~{\bf 34}, L391 (2001) and {\bf 34}, 5679 (2001).

\bibitem{r3} C.~M.~Bender, D.~C.~Brody, and H.~F.~Jones, Phys.~Rev.~Lett.~{\bf
89}, 270401 (2002).

\bibitem{r4} C.~M.~Bender, S.~Boettcher, and P.~N.~Meisinger,
J.~Math.~Phys.~{\bf 40}, 2201 (1999).
% ``$\mathcal{PT}$-Symmetric Quantum Mechanics''

\bibitem{r5} A.~Nanayakkara, Czech.~J.~Phys.~{\bf 54}, 101 (2004) and
J.~Phys.~A: Math.~Gen.~{\bf 37}, 4321 (2004).

\bibitem{r6} M.~Feigenbaum, Los Alamos Science {\bf 1}, 4 (1980).
\end{thebibliography}
\end{document}